\documentclass[twoside,onecolumn,journal,11pt]{IEEEtran}

\usepackage{graphicx}
\usepackage{amssymb}
\usepackage{amsmath}
\usepackage{rotating}
\usepackage{latexsym}

\renewcommand{\baselinestretch}{1.5}

\newcommand{\kep}{k_\text{ep}}

\newcommand{\ktrans}{K^\text{trans}}
\newcommand{\vie}{v_e}

\newcommand{\ie}{\textit{i.e.}}

\newcommand{\etal}{\textit{et~al.}}

\def\eref#1{Eqn.~\ref{#1}}
\def \mm#1{\ensuremath{\boldsymbol{#1}}}

\title{A semi-parametric technique for the quantitative analysis of Dynamic contrast-enhanced MR images based on Bayesian P-Splines}

\author{Volker J Schmid, Brandon Whitcher, Anwar R Padhani and Guang-Zhong Yang$^*$
\thanks{
Volker Schmid and Guang-Zhong Yang* are with the Institute of 
Bio\-medical Engineering, Imperial College, London SW7~2AZ, United
Kingdom.}
\thanks{Brandon Whitcher is with the Clinical Imaging Centre,
 GlaxoSmithKline, London, United Kingdom.}
\thanks{Anwar Padhani is with Paul Strickland Scanner Centre,
Mount Vernon Hospital, Northwood, United Kingdom.
}
\thanks{Asterisk
indicates corresponding author (e-mail g.z.yang@imperial.ac.uk).}
\thanks{
Preliminary results initially presented at 9th International Conference on Medical Image Computing and Computer Assisted Intervention, October 1st to 6th 2006, Copenhagen \cite{schmid06m}.}}

\begin{document}

\maketitle

\begin{abstract}
Dynamic Contrast-enhanced Magnetic Resonance Imaging (DCE-MRI) is an
important tool for detecting subtle kinetic changes in cancerous
tissue. Quantitative analysis of DCE-MRI typically involves the
convolution of an arterial input function (AIF) with a nonlinear
pharmacokinetic model of the contrast agent concentration.  Parameters
of the kinetic model are biologically meaningful, but the optimization
of the non-linear model has significant computational issues.  In
practice, convergence of the optimization algorithm is not guaranteed
and the accuracy of the model fitting may be compromised.  To overcome
this problems, this paper proposes a semi-parametric penalized spline
smoothing approach, with which the AIF is convolved with a set of
B-splines to produce a design matrix using locally adaptive smoothing
parameters based on Bayesian penalized spline models (P-Splines).  It
has been shown that kinetic parameter estimation can be obtained from
the resulting deconvolved response function, which also includes the
onset of contrast enhancement.  Detailed validation of the method,
both with simulated and {\em in vivo} data, is provided.
\end{abstract}


\section{Introduction} 

Evaluation of tissue kinetics in cancer with Dynamic Contrast-Enhanced
Magnetic Resonance Imaging (DCE-MRI) has become an important tool for
cancer diagnosis and quantification of the outcome of cancer therapies
\cite{collins04}.  For DCE-MRI, after the administration of a contrast
agent, such as \textit{Gadolinium diethylenetriaminepentaacetic acid}
(Gd-DTPA), a dynamic imaging series is acquired. Typically,
$T_1$-weighted sequences are used to assess the reduction in $T_1$
relaxation time caused by the contrast agent. The contrast agent
concentration time series can be estimated from the observed signal
intensity using proton density weighted MRI after calibration or using
multiple flip angle sequences \cite{parker03,fram87}.  With Gd-DTPA,
the agent does not enter into cells, so DCE-MRI depicts exchange
between the vascular space and the extra-vascular extra-cellular space
(EES). 

Current approaches to quantitative analysis of DCE-MRI generally rely
on non-linear pharmacokinetic models. These models are usually derived
from the solution to a system of linear differential equations, which
describe the blood flow in the tissue\cite{larsson92}.  In practice,
single-compartment models may not always be suitable, and thus more
complex models have been suggested. They include the ``extended''
Tofts--Kermode model (see \eref{onecomp}) and the tissue homogeneity
approach \cite{buckley02,stlawrence98a}.  However, non-linear
regression models are difficult to optimize and the estimation of the
parameters depends on the initial values of the algorithm
\cite{buckley02}.

Recently, model-free techniques have received extensive attention in
quantitative imaging. Neural networks are used for tissue
classification \cite{twellmann05,chuang05} and semi-parametric methods
in the kinetic modeling of dynamic PET imaging
\cite{cunningham93,gunn02}. For the quantification of first-pass
myocardial perfusion, Jerosch-Herold \etal~\cite{jeroschherold02}
proposed a model-free approach. The formulation of the response
function based on a B-spline polynomial representation, however, is
ill-conditioned and a first-order difference penalty spline (Tikhonov
regularization) has been imposed.

As an alternative, Bayesian inference has also been investigated to
replace traditional least-square fitting algorithms in MRI
\cite{goessl01a,schmid05m}. Bayesian methods allow a more accurate
description of the estimation uncertainty and can effectively reduce
bias \cite{orton07}. Furthermore, Bayesian hierarchical models can
incorporate contextual information in order to reduce estimation
errors \cite{schmid06}.  A popular, general approach for
semi-parametric modeling is based on Penalty splines, or P-splines
\cite{eilers96,marx98}.  The function under consideration is
approximated by a linear combination of a relatively large number of
B-spline basis functions.  A penalty based on $k$-th order differences
of the parameter vector ensures smoothness of the function.  For
selecting the penalty weight (or smoothing parameter), the L-curve
method \cite{johnston00} or cross validation can be used.  In Bayesian
frameworks, P-splines regression parameters can be estimated jointly
with the penalty weight and hence allow for adaptive smoothing
\cite{lang02,lang04}.

In this paper, we propose a Bayesian P-spline model to fit the
observed contrast agent concentration time curves. The model uses a
locally adaptive smoothing approach as the observed time series signal
varies rapidly in the first minute after injecting the contrast bolus.
The proposed algorithm provides a semi-parametric deconvolution
approach and results in a smooth response function, along with the
corresponding estimates of uncertainty.  Kinetic parameters are
robustly derived by fitting a non-linear model to the estimated
response function.  In addition, the exact onset of the contrast
uptake can be determined by using information derived from the
Bayesian estimation process.  Detailed validation of the method both
with simulated and \textit{in vivo} data of patients with breast
tumors is provided.

\section{Theory and Methods}

\subsection{Standard kinetic models for DCE-MRI}
\label{basic.pk}

The standard parametric model for analyzing contrast agent
concentration time curves (CTCs) in DCE-MRI is a single-compartmental
model \cite{kety60b}, where the solution can be expressed as a
convolution of the arterial input function (AIF) with a single
exponential function \cite{tofts99}, \ie,
\begin{equation}
C_t(t) =  C_p(t) \otimes \ktrans \exp (-\kep t). \label{toftskermode}
\end{equation}
Here, $C_t(\cdot)$ denotes the concentration of the contrast agent,
$C_p(\cdot)$ denotes the AIF, $\ktrans$ represents the volume transfer
constant between blood plasma and EES, and $\kep$ represents the rate
constant between EES and blood plasma.  In this study, we use an
extended version of the Tofts--Kermode model \cite{buckley02} as a
reference parametric model given by
\begin{equation}
C_t(t) = v_pC_p(t) + C_p(t) \otimes \ktrans \exp (-\kep t), \label{onecomp}
\end{equation}
where the additional parameter $v_p$ represents the fraction of
contrast agent in the vascular compartment of the tissue.

The arterial input function $C_p$ describes the input of the contrast
agent into the tissue.  DCE-MRI studies typically use a standardized
double exponential AIF given by \cite{tofts91}
\begin{equation}
C_p(t)= D \sum_{i=1}^2 a_i \exp(-m_i t) \label{aif}
\end{equation}
with values $a_1=24$ kg/l, $a_2=6.2$ kg/l, $m_1=3.00$ min$^{-1}$ and
$m_2=0.016$ min$^{-1}$ \cite{fritzhansen96}. In \eref{aif}, $D$ is the
actual dosage of tracer in mmol/kg.  With this explicit form of the
AIF, the convolution in \eref{onecomp} can be derived analytically,
\ie,
\begin{equation}
  C_t(t)
 =
 v_{p} D \sum_{i=1}^2 a_i \exp(-m_i t) 
+ D \ktrans \sum_{i=1}^2 \frac{a_i \{\exp(-m_it) -
    \exp[-\kep t] \}}{\kep-m_i}. 
\label{nonlinearfunc}
\end{equation}
The kinetic parameters $\ktrans$, $\kep$ and $v_p$ are estimated by
fitting \eref{nonlinearfunc} to the observed data.  Optimization is
usually performed using the Levenberg--Marquardt algorithm
\cite{more78}.

\subsection{Bayesian P-Spline model for DCE-MRI}

Parametric models, however, may not give an accurate fit of the
observed data, as they are too simplistic and can overlook effects
such as flow heterogeneity or the water exchange effect
\cite{kroll96,landis00}.  More complex models have since been
developed \cite{stlawrence98a}, but they are more difficult to
optimize.  It has been found that semi-parametric models can fit the
data more accurately and in this study penalty splines (P-Splines) in
a Bayesian hierarchical framework are used.

\subsubsection{Discrete Deconvolution}

Mathematically, a more general expression of \eref{toftskermode} can
be written as:
\begin{equation}
C_t(t) = C_p(t) \otimes f(t), \label{general}
\end{equation}
where $f(t)$ is the response function in the tissue.  Assuming that
$C_p(\cdot)$ and $f(\cdot)$ are constant over small intervals $\Delta
t$, a discretized form of \eref{general} is given by
\begin{equation}
  C_t(\tau_i) = \sum_{j=1}^T C_p(\tau_i-t_j) \, f(t_j)
  \Delta\,t=\sum_{j=1}^T A_{ij}f(t_j), \label{discretematrixmodel}
\end{equation}
where $C_t$ is measured on discrete time points
$\tau_1\,\ldots,\tau_n$.  The $n{\times}T$ matrix \mm{A} may be
interpreted as a convolution operator and is defined by
\begin{equation}
  A_{ij} = \left\{\begin{array}{ll}
  C_p(t_{n_i-j+1})\Delta~t \qquad & \text{if $\tau_i \le t_j$};\\
  0 &  \text{otherwise},
  \end{array}\right.
\end{equation}
where $n_i$ is the maximum index $j$ for which $\tau_i\le t_j$ holds.
It is worth noting that the input function $C_p(t)$ is measured -- or
evaluated from \eref{aif} -- at time points $t_1,\ldots,t_T$, which
can be different from $\tau_1,\ldots,\tau_n$.

\subsubsection{Penalty Splines}

By solving \eref{discretematrixmodel}, the response function $f(t)$
can be deconvolved from $C_p(t)$. However, this system may be
numerically unstable, \ie, the deconvolved response function is
susceptible to noise.  To overcome this problem, we assume that $f(t)$
is a smooth, $k$-times differentiable function.  To this end, a
B-spline representation of the response function is used in this
study, \ie,
\begin{equation}
  f(t)=\sum_{j=1}^p \beta_j B_{tj}, \label{responsefx}
\end{equation}
where $\mm{B}$ is the $n{\times}p$ design matrix of $k$th order
B-splines with knots $s_1,\ldots,s_{p+k}$. In vector notation
$\mm{f}=(f(t_1),\ldots,f(t_T))'$ and \eref{responsefx} may be
expressed as
\begin{equation}
\mm{f}=\mm{B}\mm{\beta}.  \label{nonparflowmodel}
\end{equation}
Accordingly, \eref{discretematrixmodel} can be written as
\begin{equation}
  \mm{C_t} = \mm{A}\mm{f} = \mm{A}\mm{B}\mm{\beta} = \mm{D}\mm{\beta},
  \label{endmodel}
\end{equation}
where $\mm{D}=\mm{A}\mm{B}$ is a $n{\times}p$ design matrix,
representing the (discrete) convolution of the AIF with the B-spline
polynomials.

To enhance the numerical stability, a penalty on the B-spline
regression parameters is introduced, such that
\begin{equation}
  \beta_t = 2\beta_{t-1} - \beta_{t-2} + e_t \qquad 
  \text{for $t=3,\ldots,p$}. \label{rw}
\end{equation}
These models are known as penalty splines or P-Splines
\cite{eilers96,marx98} as they penalize the roughness of the function
$f(t)$, and therefore act as a denoising method. As $\mm{C_t}$
exhibits a sharp initial increase followed by a sharp decrease at the
beginning of the dynamic series, the penalty has to be locally
adaptive.  To this end, we use a Bayesian hierarchical framework for
parameter inference.

\subsubsection{Bayesian hierarchical framework}

In Bayesian inference, \textit{a priori} information, \ie, information
available before observation of measurement, has to be expressed in
terms of probability distributions. Here, our prior knowledge is the
assumption about the model.  In this paper, we assume that the
observed contrast agent concentration $\mm{C_t}$ is noisy realization
of the true model (\eref{endmodel}), \ie,
\begin{equation}
  C_t(t) \sim \text{N}(\mm{D}_{t}\mm{\beta},\sigma^2) \quad \text{for all $t$}, \label{data.model}
\end{equation}
where $\mm{D}_t$ denotes the $t$th row of $\mm{D}$. That is, \textit{a
priori} the error is assumed to be Gaussian distributed with unknown
variance $\sigma^2$. We use a relatively flat prior for the variance
parameter, \ie,
\begin{equation}
\sigma^2 \sim \mbox{IG}(1,10^{-5}),\label{sigma}
\end{equation}
where $\text{IG}$ denotes the Inverse Gamma distribution. 

The penalty in \eref{rw} can be expressed as \textit{a priori}
distribution on $\beta$ \cite{lang04},
\begin{equation}
  \beta_t \sim \text{N}(2\beta_{t-1}-\beta_{t-2},\delta_t^2) \qquad 
  \text{for $t=3,\ldots,p$}, \label{rw2}
\end{equation}
where $\delta_t^2$ is the variance of $e_t$. For a locally adaptive
estimation of $\delta_t^2$, the following prior model is used
\begin{equation}
\delta_i^2 \sim \text{IG}(10^{-5},10^{-5}) \qquad \text{ for $i=3,\ldots,p$},\label{delta}
\end{equation}
which allows time varying smoothness penalties.

\subsubsection{Evaluation of the posterior distribution}
\label{fullcond} 

Bayes' theorem was used to calculate the posterior distribution of the
parameter vector which is, up to a normalizing constant, given by
\begin{equation}
p(\mm{\beta},\mm{\delta^2},\mm{\sigma^2})\propto \ell(\mm{C_t}|\mm{\beta},\mm{\sigma^2})p(\mm{\sigma^2})p(\mm{\beta}|\mm{\delta^2})p(\mm{\delta^2}).\label{e16}
\end{equation}
A closed-form solution of \eref{e16} is not possible, and thus, Markov
chain Monte Carlo (MCMC) techniques have been used to assess the
posterior distribution \cite{gilks96}.  The full conditional of
$\mm{\beta}$, \ie, the joint distribution of $\mm{\beta}$ given all
other parameters and the data, $\mm{C_t}$, is a $p$-dimensional
multivariate normal distribution
\begin{equation}
\mm{\beta}|\mm{C_t},\mm{\delta^2},\mm{\sigma^2} \sim \mbox{\bf N}_p\left(
\mm{C_t}'\mm{D}(\mm{D}'\mm{D}+\mm{R})^{-1},
\sigma^2(\mm{D}'\mm{D}+\mm{R})^{-1}\right),\label{post:beta}
\end{equation}
where $\mm{R}$ is the inverse covariance matrix of the prior
distribution of $\mm{\beta}$ \cite{lang04}.
The full conditionals of both variance parameters are independent
Inverse Gamma distributions
\begin{eqnarray}
\delta^2_t|\mm{\beta} &
 \stackrel{iid}\sim &
 \mbox{IG}\left(10^{-5}+0.5, \label{post:delta}
 \, 10^{-5}+0.5(\beta_t-2\beta_{t-1}+\beta_{t-2})\right) \text{ \quad for } t=3,\ldots,p, \\
\sigma^2|\mm{\beta},\mm{C_t} & \sim & \mbox{IG}\left(1+\frac{T}{2},
 \, 10^{-5}+0.5(C_t-\mm{D_t}\mm{\beta})'(C_t-\mm{D_t}\mm{\beta})\right).\label{post:sigma}
\end{eqnarray}
To assess the posterior distribution, samples of the parameters
$\mm{\beta},\sigma^2$ and $\mm{\delta}^2$ are drawn alternately from
\eref{post:beta}, \eref{post:delta} and \eref{post:sigma}. After a
sufficient burn-in period, the MCMC algorithm produces samples from
the posterior distribution.

\subsection{Estimating kinetic parameters from the semi-parametric technique}
\label{sec:est}

The advantage of parametric models defined by Eqns.~\ref{toftskermode}
and \ref{onecomp} is that they contain biologically meaningful and
interpretable parameters. The semi-parametric technique only provides
a fit to the CTC and a de-convolved and de-noised response function,
but not kinetic parameters. In this section, we introduce methods for
estimating kinetic parameters and the onset of the contrast uptake
from the estimated response function. We make use of the fact that the
Bayesian approach provides a rich source of information via the
posterior distribution of the response function.

\subsubsection{Determining the onset of contrast uptake}

In practice, the time delay between the injection of the contrast
agent and the arrival of the tracer in the tissue of interest is
unknown. However, it is important to correctly estimate the delay for
a robust estimation of the kinetic parameters \cite{jeroschherold04}.
The Bayesian approach yields information on the uncertainty for each
parameter in the model and for all transformations of the parameters
such as the estimated contrast agent concentration
$\mm{C_t}=\mm{D}\mm{\beta}$. This results in a point wise Credible
Interval (CI) for the contrast agent concentration.  From this, we can
compute the minimal time $t^*$, where the 99\% CI does not cover
0. That is, at time $t^*$ we are 99\% confident that the contrast
agent has already entered the tissue.

Assuming that the initial slope of the contrast agent concentration
time curve is approximately linear, we can compute the onset time by
drawing a line from $C_t(t^*)$ to 0 with the gradient $dC_t(t^*)/dt$.
The first order derivative of $C_t(t^*)$ can be computed by
\[ \frac{d}{dt} C_t(t) = \frac{d}{dt}\left[C_p(t)\otimes f(t)\right] =
C_p(t)\otimes \frac{d}{dt}f(t).\] Since $f(t)$ is a spline, the
derivative can be computed as \cite{tsao93}
\[
 f(t)=\sum_{j=1}^{n-k}\gamma_j B_{t(j+1)}^{(k-1)},
 \]
where $\mm{B^{(k-1)}}$ is the design matrix of $(k-1)$th order
B-Splines and $\gamma_j$ is defined via
\[ \gamma_j = \frac{k}{s_{j+k+1}-s_{j+1}}(\beta_{j+1}-\beta_{j}) \text{ for } j=1,\ldots,J-1.\]
In this paper, we propose the following algorithm to estimate the
onset of the contrast uptake:
\begin{enumerate}
\item Find the minimum time $t^*$, for which the contrast concentration 
significantly exceeds zero,
\item Compute the gradient of the estimated CTC at $t^*$,
\item Calculate the enhancement onset time as
$t_0=t^*-\frac{C_t(t^*)}{dC_t(t^*)/dt}$.\\
\end{enumerate}
DCE-MRI studies typically assume that the onset of the enhancement is
the same over the whole region of interest (ROI). In this case, the
median of the estimated $t_0$ for all voxels in the ROI may be used as
an estimate of the global value. However, for larger ROIs, local
estimates of the onset may be required, and can be computed from the
proposed technique.

\subsubsection{Obtaining kinetic parameters}
\label{OKP}
 
The semi-parametric technique provides a de-convolved and de-noised
response function. In order to obtain kinetic parameters, we can fit a
non-linear function to the estimated response. To this end, the
following model has been used:
\begin{equation}
  f(t)  =  F_p 
\cdot\left\{\begin{array}{ll}
E  \exp[-(t-t_0-T_c)EF_p/\vie] & \text{for $t \ge (T_c+t_0)$},\\
  1 & \text{for $t_0\le t< (T_c+t_0)$},\\
0 & \text{for $t< t_0$},\\
  \end{array}\right. \label{rm}
\end{equation}
where $T_c = v_p/\ktrans$ is the transit time through the capillary,
$\vie = \ktrans/\kep$ is the volume fraction of EES, $E$ is the
extraction fraction, and $F_p$ is the mean plasma flow. This model is
similar to the adiabatic approximation of tissue homogeneity (AATH)
\cite{stlawrence98a}. In this model extraction fraction $E$ and mean
plasma flow $F_p$ may be not identifiable, but the product
$EF_p=\ktrans$ is.

It should be noted that the Bayesian methodology provides not just one
response function, but a (posterior) distribution of response
functions.  In order to obtain a distribution of the estimated
parameters, the model in \eref{rm} can be fitted to each response
function in the sample using the Levenberg--Marquardt optimization
algorithm.  The median of the posterior distributions may be used to
estimate the parameters. The estimation error can be computed using
the standard errors across the samples and intervals can be
constructed from quantiles of the posterior distributions.

\subsection{Data Collection}

Simulated DCE-MRI data of CTCs with known kinetic parameters were
previously published in \cite{buckley02}.  We use the first series
which was designed to be representative of data acquired from a breast
tumor.  Data was simulated using MMID4, part of a software made
available by the National Simulation Resource, Dept.~of
Bioengineering, University of Washington
(\textit{http://www.nsr.bioeng.washington.edu}).  Estimates from the
literature were used as baseline values.  For twelve further
experiments, one of the kinetic parameters $F_p$, $v_p$ and $PS$ was
changed four times while the other two where held fixed at the
baseline values (see Tab.~\ref{tab:simvalues}). Data was sampled at 1
Hz.

For a second set of simulated data, a time lag of up to 30 seconds was
added to the simulated data to evaluate the effect of lagged contrast
uptake.  To make the experiment more realistic for DCE-MRI, data was
down sampled to 1/8 Hz; scans in DCE-MRI experiments are typically
acquired every 4--12 seconds.

\textit{In vivo} data was derived from twelve patients with primary
breast cancer (median age 46 years; range 29-70). Each patient was
scanned twice, once before and once after two cycles of chemotherapy.
Scans were performed with a 1.5~T Siemens MAGNETOM Symphony scanner
(TR = 11~ms and TE = 4.7~ms; 40 scans with four sequential slices were
acquired in about 8~minutes).  A dose of D = 0.1~mmol/kg body weight
Gd-DTPA was injected at the start of the fifth acquisition using a
power injector.  This study was provided by the Paul Strickland
Scanner Centre at Mount Vernon Hospital, Northwood, UK.  Data from
this study was acquired in accordance with the recommendation given by
Leach et al.\cite{leach05} and previously reported
\cite{ahsee04,schmid05m}.
Tumor ROIs were drawn by an expert radiologist based on the
dynamic $T_1$ images.

\section{Results}

\begin{figure}[tb]
\begin{center}
  \includegraphics*[width=\textwidth]{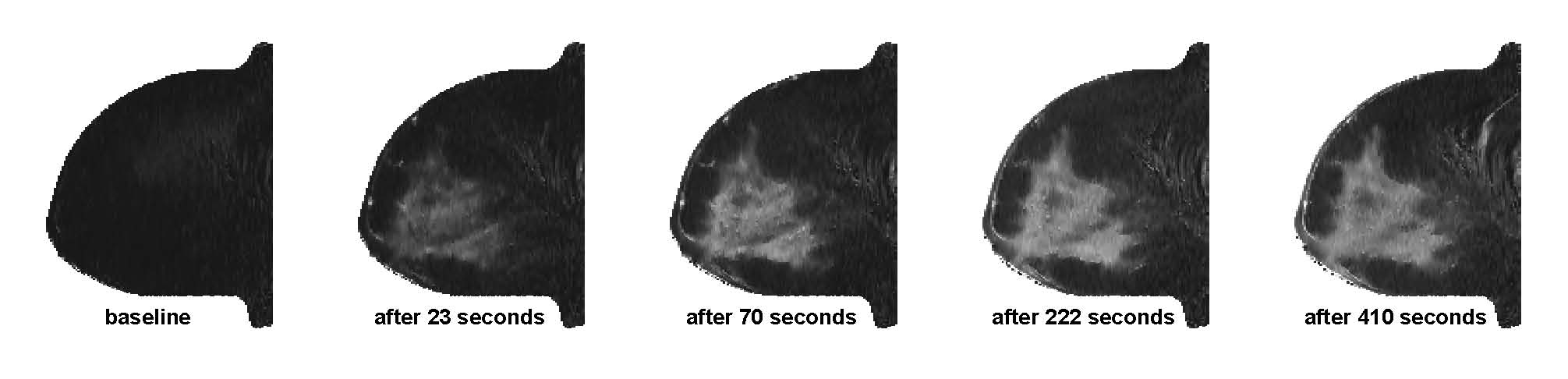}
  \caption{Time series of contrast concentration images at baseline
    and after 23, 70, 222 and 410 seconds - central slice from first
    scan of patient 9.\label{fig:band}}
\end{center}
\end{figure}

To validate the proposed methods, the first simulation study evaluates
the fit of the semi-parametric technique to simulated data in
comparison to a parametric method.  The second simulation study
evaluates the estimation algorithm for the onset of the enhancement
and the kinetic parameters when the arrival of tracer in the tissue
ROI is lagged.  For \textit{in-vivo} validation, clinical data of 24
DCE-MRI scans of breast cancer patients were analyzed. A series of
dynamic images from a patient is depicted in Fig.~\ref{fig:band}.

\subsection{Simulation studies}


The semi-parametric technique provides an accurate fit to the observed
contrast agent concentration time curve. The sum of the squared
residuals (SSR) for the simulated data has a range of $1.89\cdot
10^{-4}$ to $2.13\cdot 10^{-3}$ with a mean of $1.13\cdot 10^{-3}$
over all the 13 experiments. With the reference parametric model, the
SSR has a range of $0.198$ to $1.520$ with a mean of $0.661$, \ie, the
fit to the observed data was poor in the reference parametric model
compared to the proposed semi-parametric technique.

\begin{figure}[tb]
\begin{center}
\begin{minipage}[]{0.32\textwidth}
  \centering
  \includegraphics*[width=0.95\textwidth]{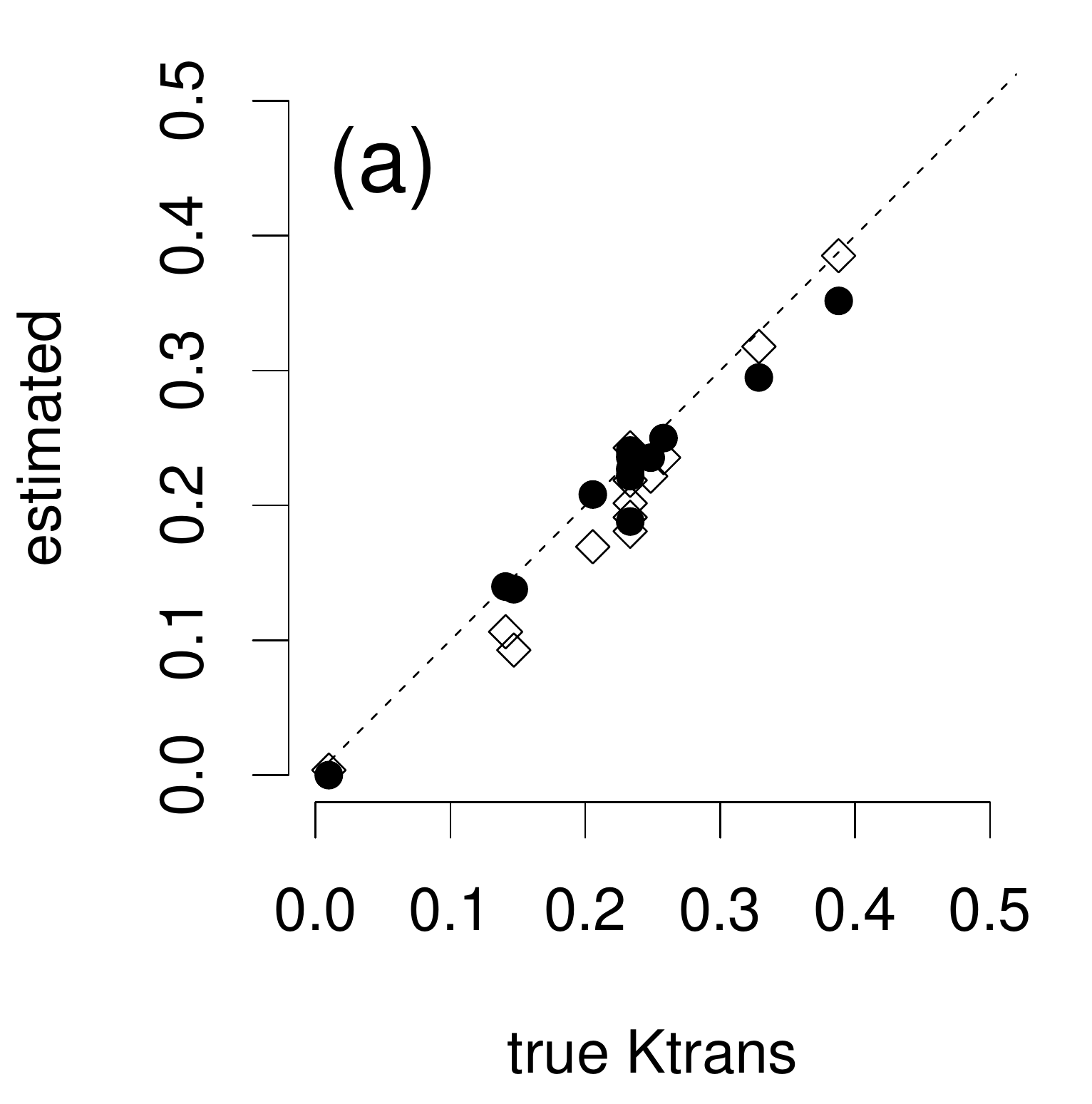}
\end{minipage}
\begin{minipage}[]{0.32\textwidth}
  \centering
  \includegraphics*[width=0.95\textwidth]{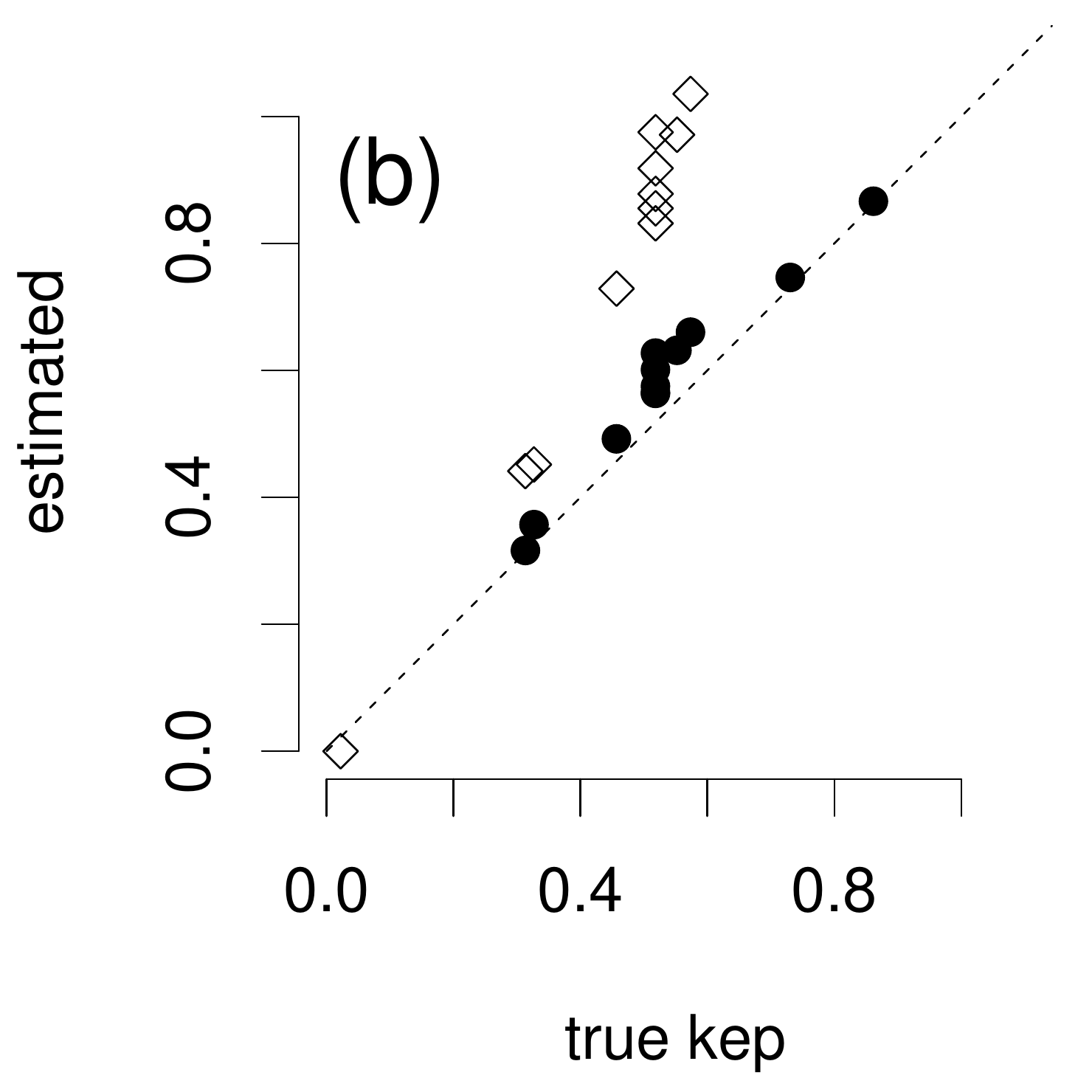}
\end{minipage}
\begin{minipage}[]{0.32\textwidth}
  \centering
  \includegraphics*[width=0.95\textwidth]{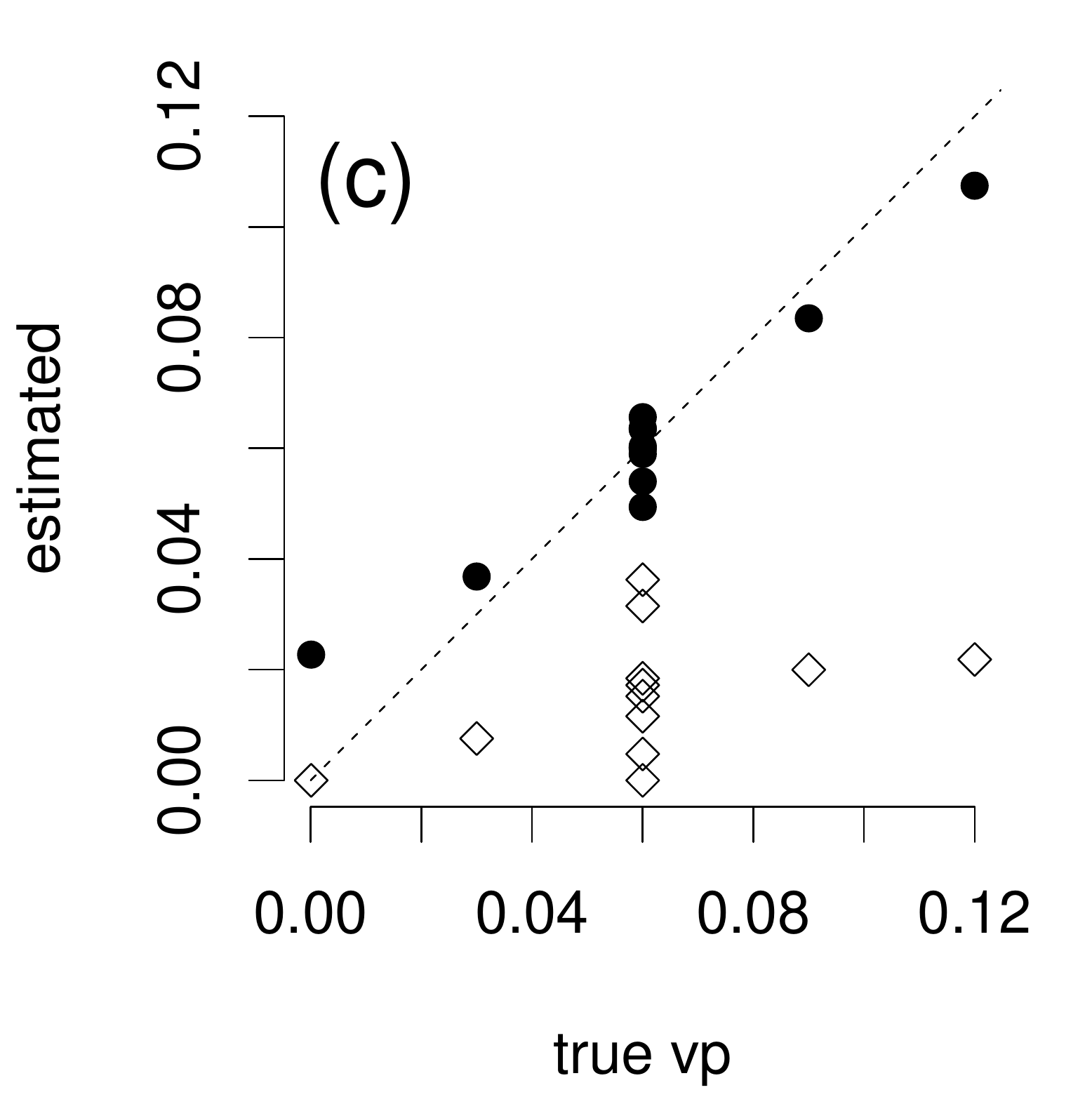}
\end{minipage}
\end{center}
\caption{Scatter plots of parameter estimates vs.~true values -- These
  figures are similar to Fig.~3 in \cite{buckley02}. Estimates of the
  parameters (a) $\ktrans$, (b) $\kep$ and (c) $v_p$. Bullets
  represent the results using the semi-parametric method, diamonds the
  results using the parametric technique.\label{fig:vgl}}
\end{figure}

Fig.~\ref{fig:vgl} depicts the the kinetic parameter estimates for the
semi-parametric technique and the reference parametric method compared
to the ground truth. Fig.~\ref{fig:vgl} (a) shows that the parameter
$\ktrans$ is underestimated with the reference parametric model -- on
average by 17.7\%, which is consistent with previously published
results \cite{buckley02}.  By contrast, $\ktrans$ estimates with the
semi-parametric technique are much closer to the ground truth. The
mean deviation from the ground truth is 6.2\%.
These results suggest that the semi-parametric technique is more
stable compared to the parametric model.
changes in $v_p$.

With the proposed semi-parametric technique, the parameters $\kep$ and
$v_p$ are also estimated accurately, but $\kep$ is slightly
overestimated by 1.5\% to 6.2\% compared to a strong overestimation of
39\% to 100\% with the reference parametric method.  For most
experiments, the $v_p$ parameter obtained from the semi-parametric
technique is much closer to the ground truth than that of the the
reference model.  For small values of $v_p$ (experiment 6), however,
the semi-parametric method shows a larger deviation from ground
truth. An important advantage of the proposed Bayesian technique is
that it not only produces point estimates, but also estimations of the
errors or interval estimators.  Tab.~\ref{fig:ktranserrorbar} gives
95\% Credible Intervals (CI) for $\ktrans$ for all the 13 experiments
conducted.  For experiments 8 and 9, where $v_p$ has larger values,
the 95\% CI are broad, but still cover the true value.  Similar CIs
are available for the other parameters. The Bayesian technique
provides important information about both the accuracy and precision
of its estimates.



To validate the proposed algorithm for estimating the onset of the
contrast uptake, a time lag of up to 30 seconds was added to the
down-sampled simulated data.  Data was analyzed with the proposed
semi-parametric technique and the enhancement onset time and the
kinetic parameters were derived by equations described in section
\ref{sec:est}.  The estimation of $t_0$ with the proposed method is
shown to be accurate.  Correlation between the estimated onset time
and ground truth is $0.9984$.  The mean difference between the true
and estimated onset time is $0.1800$ seconds with a standard deviation
of $0.9339$ seconds and a maximum absolute difference of $2.2992$
seconds.

\begin{figure}[!tb]
  \begin{center}
    \includegraphics*[width=.65\textwidth]{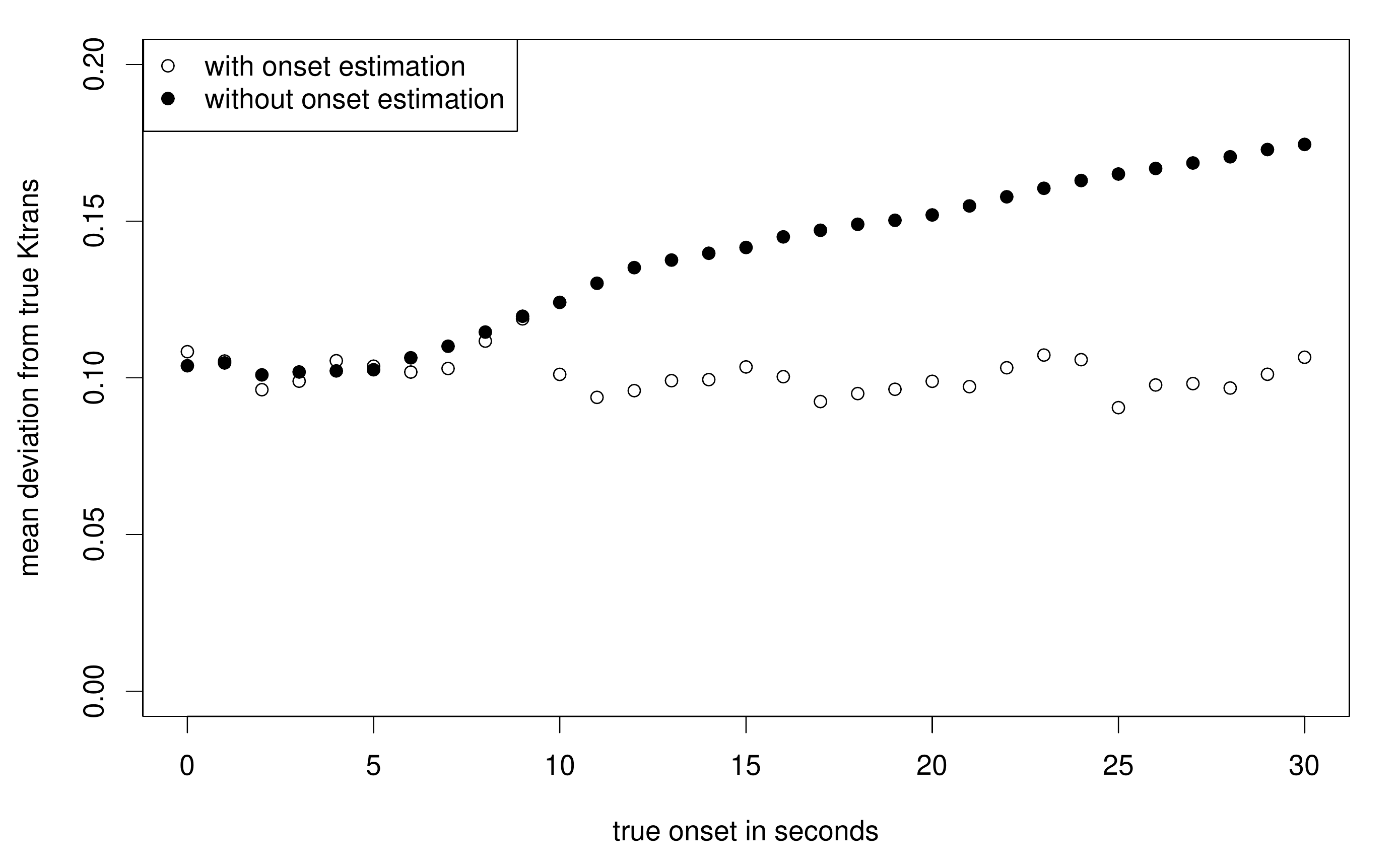}
    \caption{Mean absolute difference between true and estimated
      $\ktrans$ values plotted against true enhancement onset time in
      seconds. Solid dots are results from semi-parametric analysis
      without considering onset, circles are results from the proposed
      algorithm with onset estimation.\label{fig:sim-onset-ktrans}}
  \end{center}
\end{figure}

Fig.~\ref{fig:sim-onset-ktrans} shows the Mean Absolute Difference
(MAD) between the estimated values of $\ktrans$ and the ground
truth. The MAD is stable with increasing time lag, but changes
periodically with respect to the sampling interval of 8 seconds.  For
comparison, the lagged data was further analyzed without $t_0$
estimation.  For a small time lag up to 8 seconds, \ie, up to one
sampling interval, a lag of the onset time has negligible influence on
$\ktrans$ estimation.  However, for larger onset lags MAD increases
significantly when onset is not taken into account.

\subsection{In vivo validation}
\label{invivo}

\begin{figure}[!tb]
  \begin{center}
\begin{minipage}[]{0.45\textwidth}
  \centering
  \includegraphics*[width=0.95\textwidth]{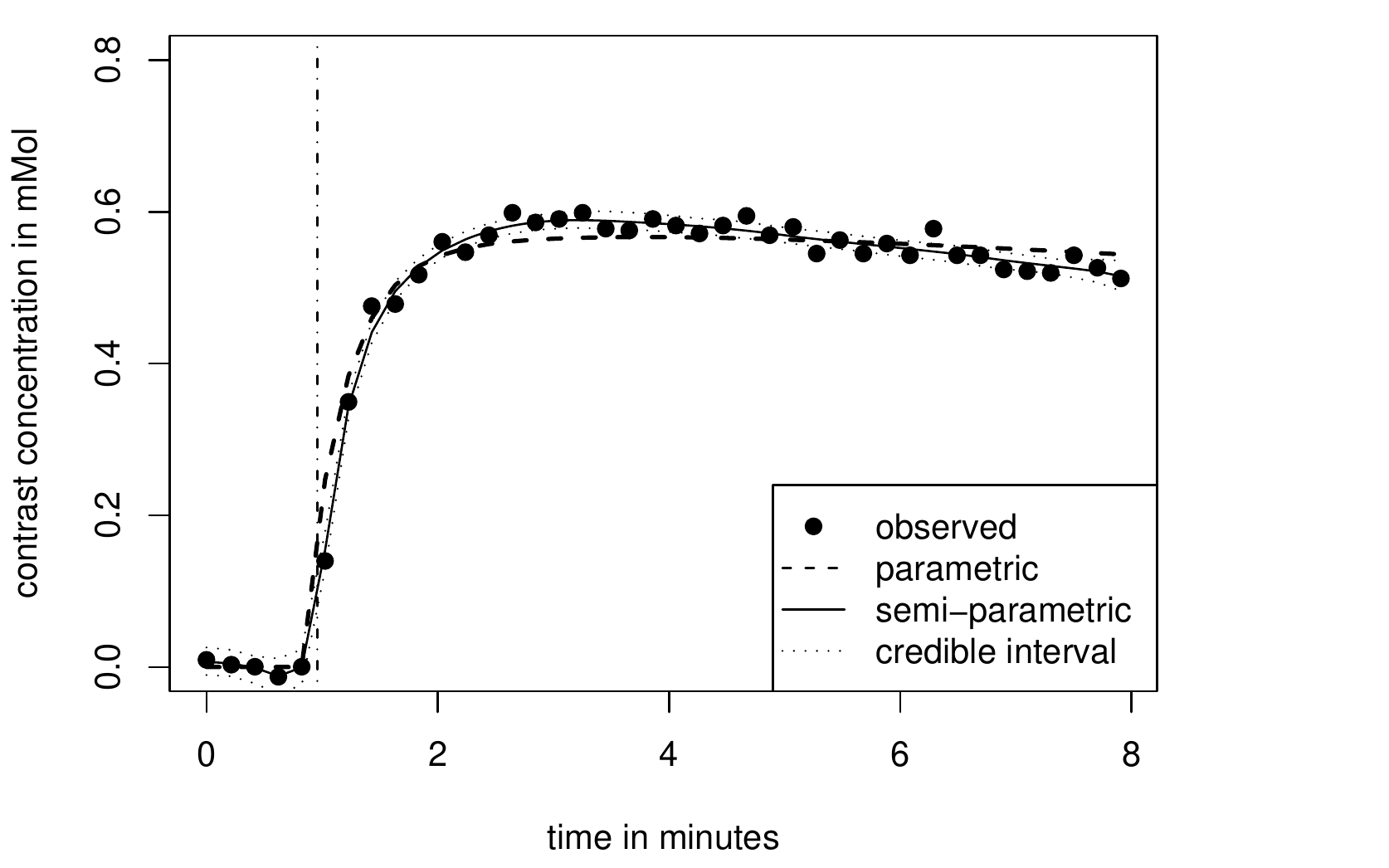}
\end{minipage}
\begin{minipage}[]{0.45\textwidth}
  \centering
  \includegraphics*[width=0.95\textwidth]{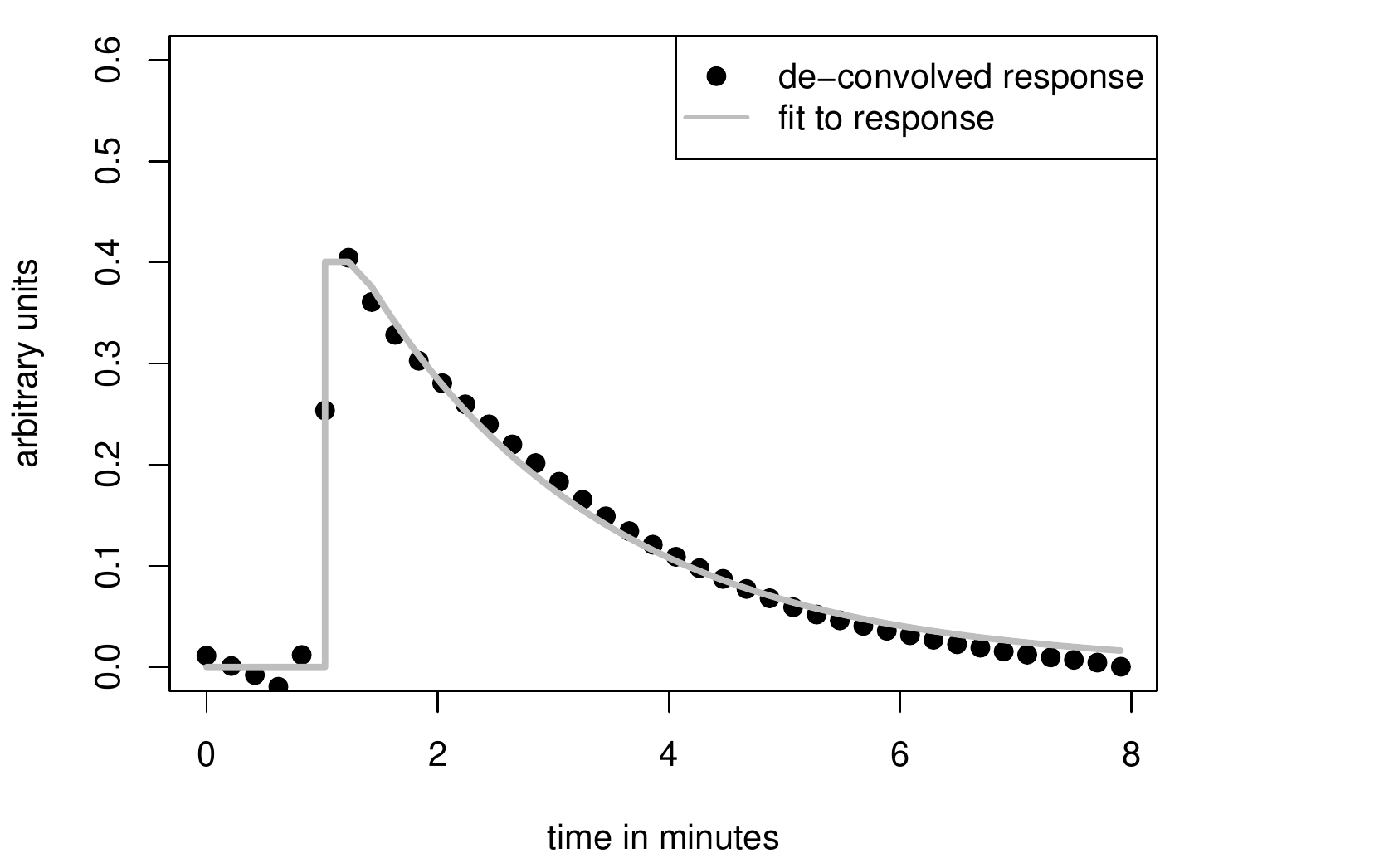}
\end{minipage}
\begin{minipage}[]{0.45\textwidth}
  \centering
  \includegraphics*[width=0.95\textwidth]{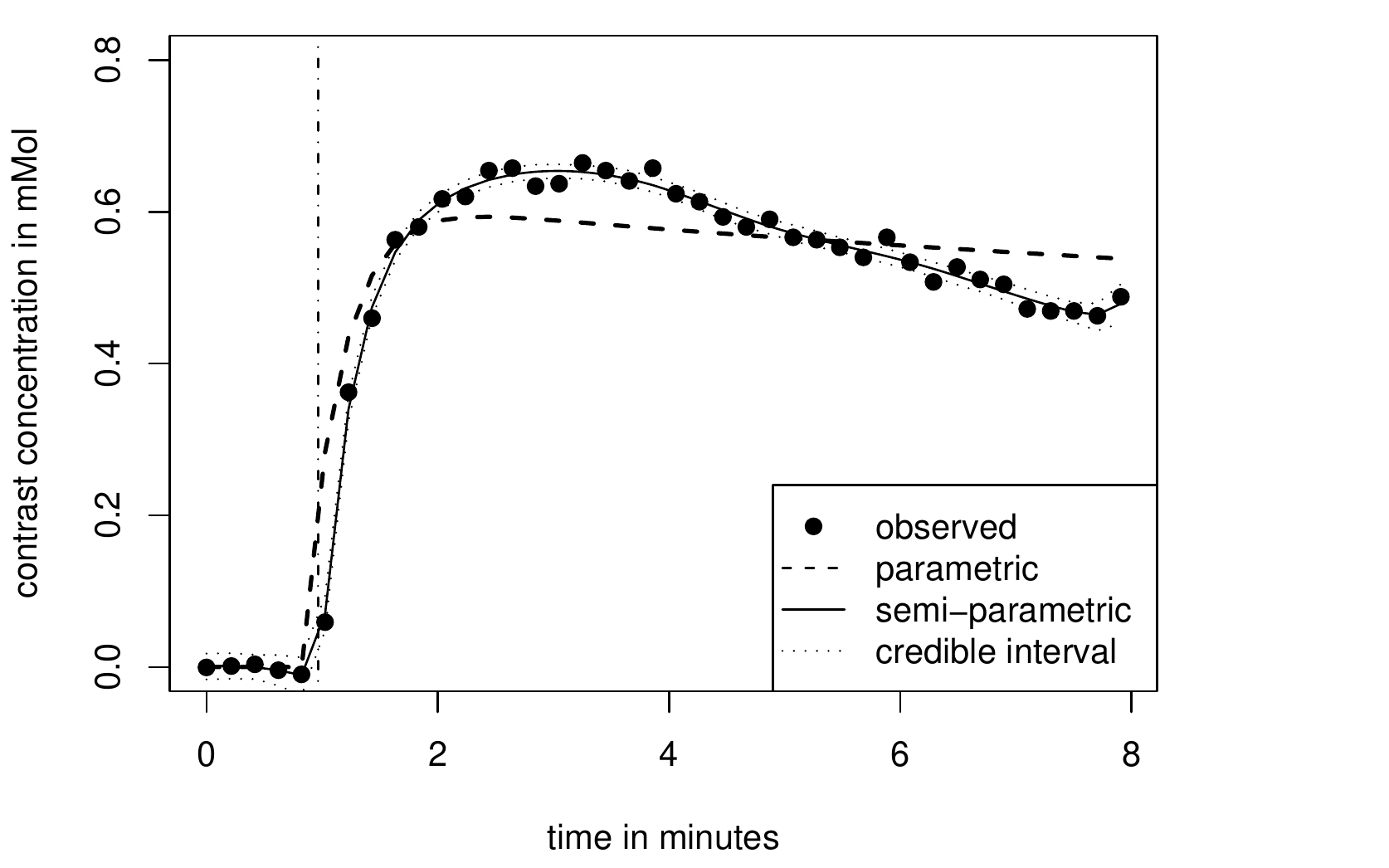}
\end{minipage}
\begin{minipage}[]{0.45\textwidth}
  \centering
  \includegraphics*[width=0.95\textwidth]{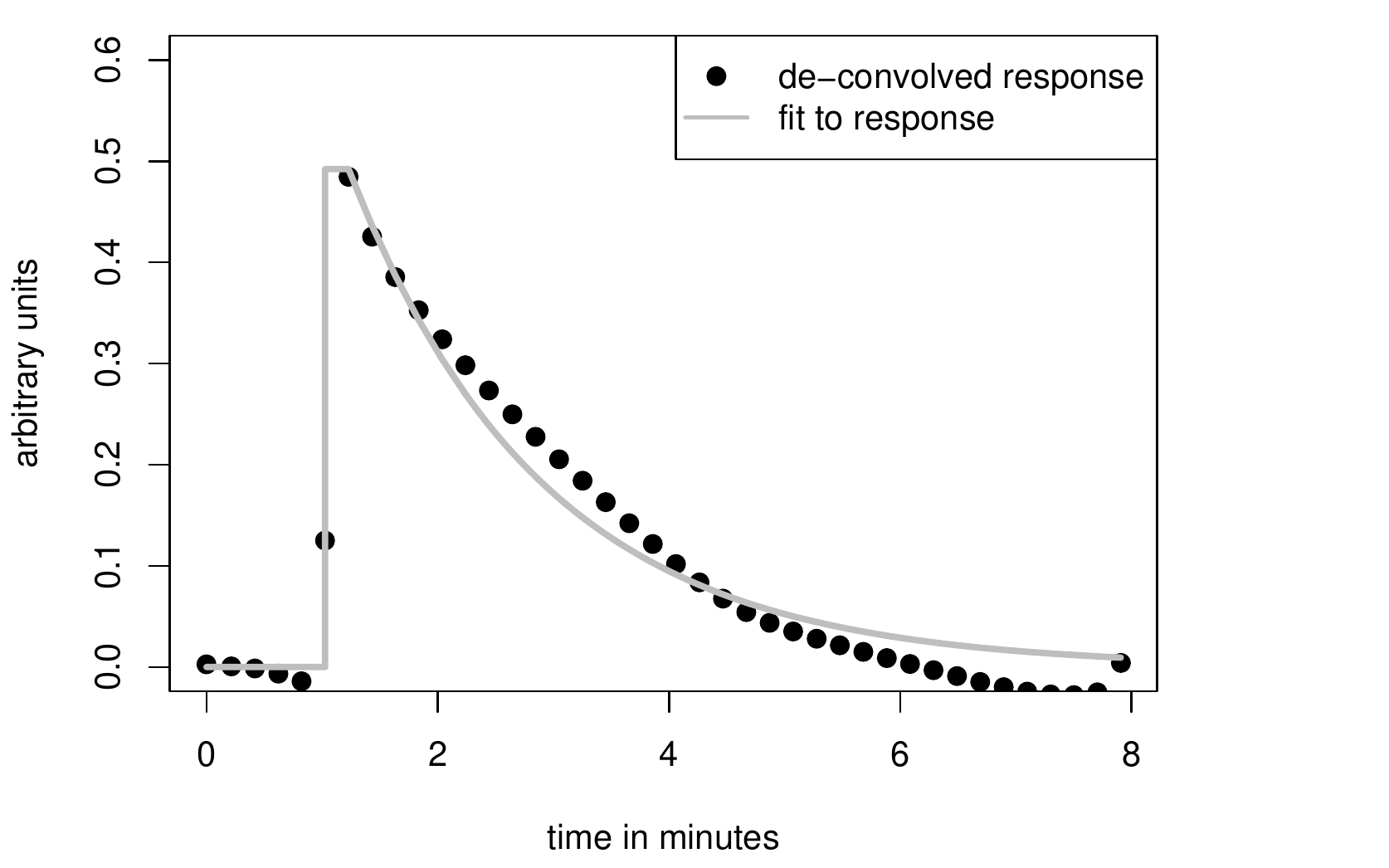}
\end{minipage}
\begin{minipage}[]{0.45\textwidth}
  \centering
  \includegraphics*[width=0.95\textwidth]{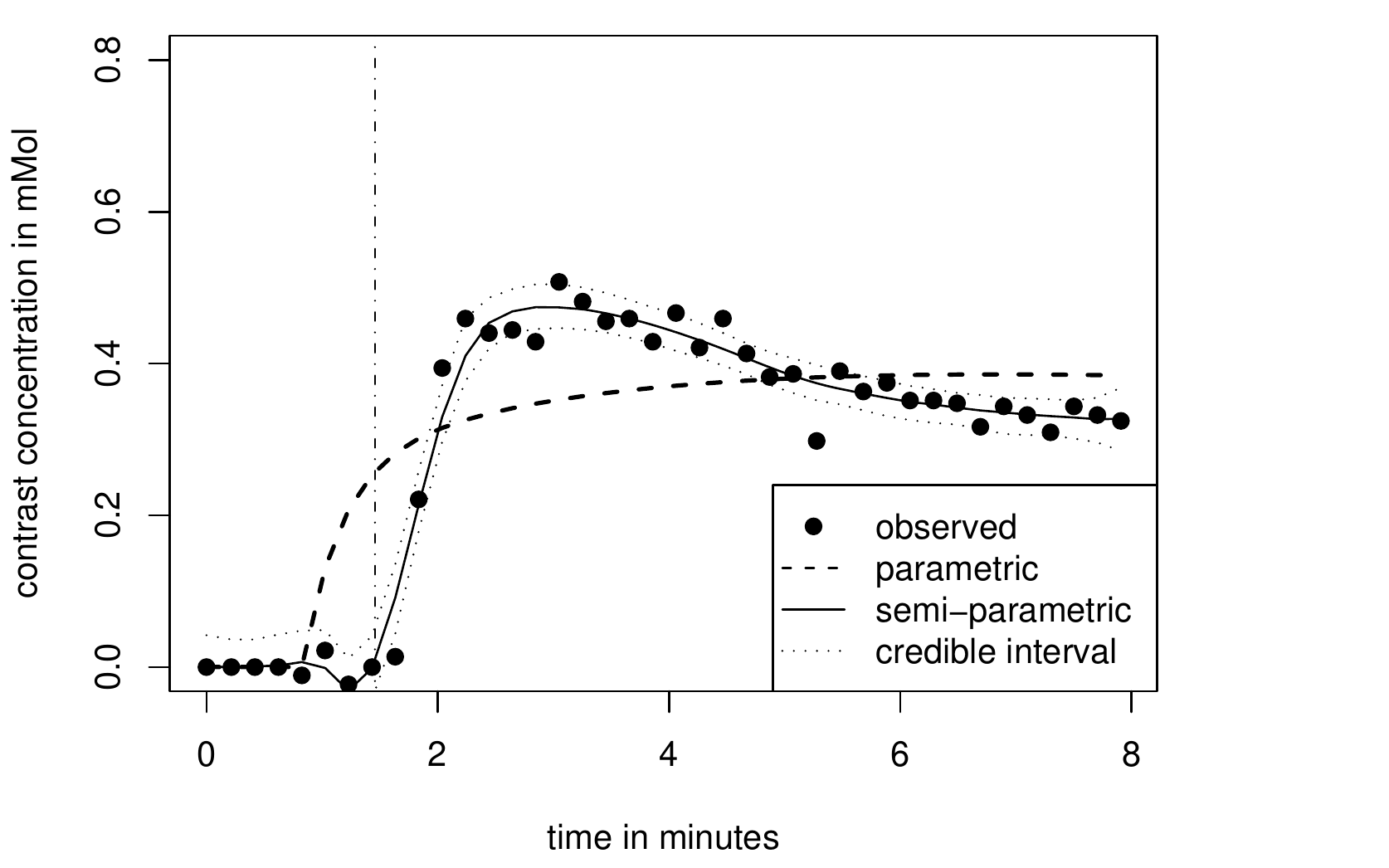}
\end{minipage}
\begin{minipage}[]{0.45\textwidth}
  \centering
  \includegraphics*[width=0.95\textwidth]{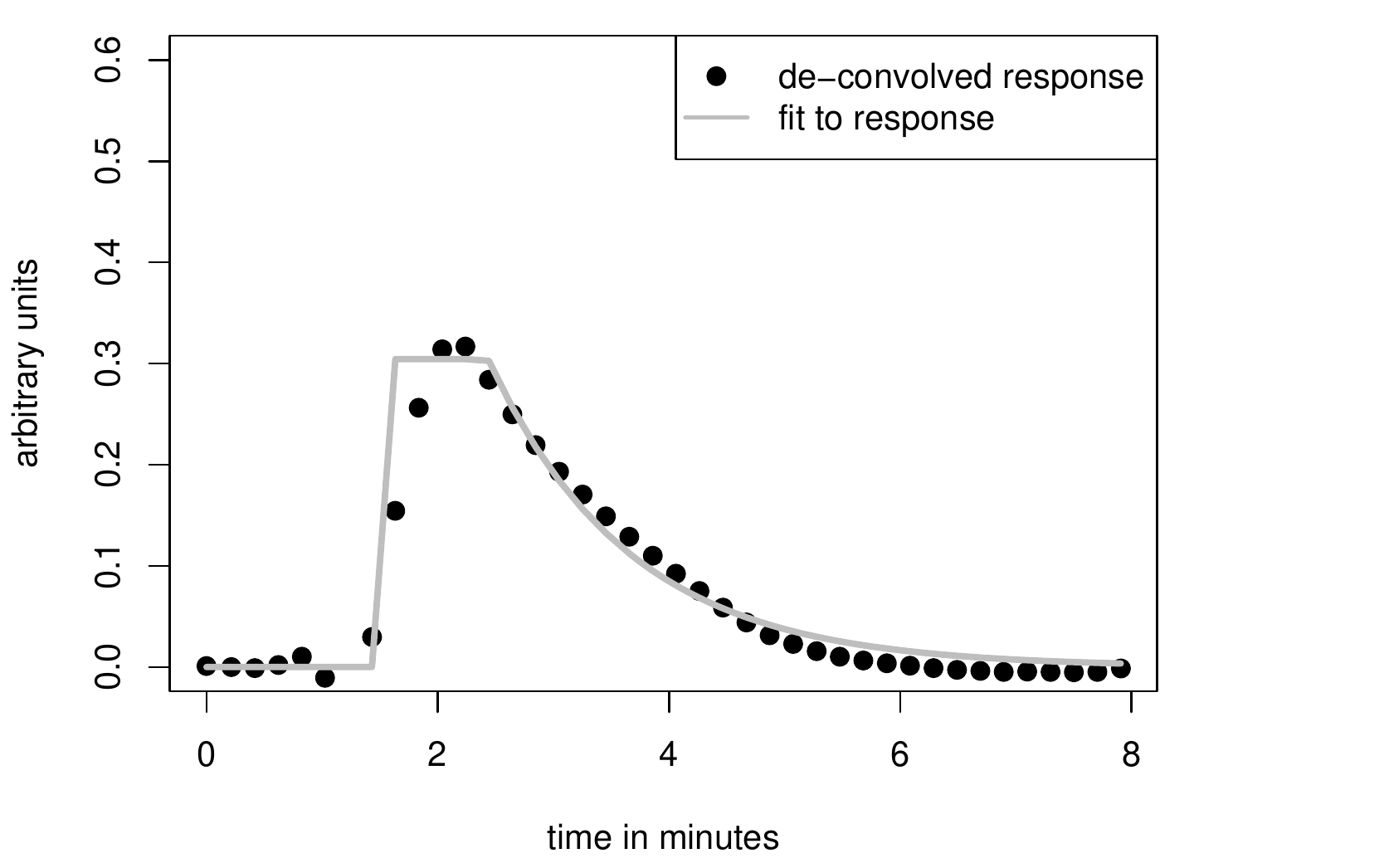}
\end{minipage}
\caption{Left column: Observed contrast agent concentration time
  series (dots) of a voxel in the ROI of the pre-treatment scans for
  three different subjects (from top to bottom: subjects 5, 9 and 4)
  along with fit by parametric (slashed line) and semi-parametric
  technique; for the semi-parametric technique the median (solid line)
  and the 95\% CI (dotted lines) are depicted.  The estimated
  enhancement onset time is marked as vertical line.  Right column:
  Median estimated de-convolved response function (dots) with fit to
  the model \eref{rm} (grey line) for the same voxels.\label{fig:fit}}
\end{center}
\end{figure}

Table \ref{tab:vglpatients} shows the sum of squared residuals (SSR),
\ie, the goodness of fit for the data from all 24 \textit{in-vivo}
scans obtained from the semi-parametric and the parametric approach.
As with the simulated data, the fit is clearly better for the
semi-parametric technique.  Table \ref{tab:t0} provides the estimated
enhancement onset time for all scans; the scanning protocol indicates
that the contrast agent was injected at the start of the fifth image,
\ie~after $49.40s$. For three subjects, the onset time is displayed in
Fig.~\ref{fig:fit} as vertical dashed line. Visual inspection of this
figure and of the other subjects shows that the estimated onset time
is consistent with the observed contrast concentration time
series. For most scans, the estimated onset time is between the start
of the fifth acquisition and the start of the sixth acquisition, which
is reasonable given the scanning protocol. For some scans, however,
the onset time is much larger due to deviation from the scanning
protocol.

Figs.~\ref{fig:fit} depicts the observed contrast concentration time
series for three voxels from three different scans together with the
estimates from the parametric reference method and those from the
proposed semi-parametric approach with 95\% CI.  It is evident that in
all voxels the semi-parametric method fits the data better than the
parametric approach. The figures also indicate that the standard
parametric model is not always appropriate for the observed data,
where the initial upslope is difficult to follow by using the
parametric model. The model also fails when the assumed onset time
strongly deviates from the observed onset time, for example for the
first scan of subject 4 (see bottom row of Fig.~\ref{fig:fit}).

In this study, the 95\% CI of the contrast agent concentration time
series nicely covers the observed CTC. About half the observed points,
however, lie outside the CI. The estimated observation error is larger
than the estimated error of the CTC, \ie, the uncertainty of the model
about the true CTC.  For the first scan of subject 4, the CI is
noticeably wider than that of other scans, \ie, the quality of the
data from this scan is lower than the quality of the other
scan. Typically, the CI is wider after the upslope of the CTC, and
narrow during wash-out. It widens again at the end of the time series,
because the spline fit at the end point relies on less data.

Fig.~\ref{fig:fit} also depicts the median of the estimated
de-convolved response function and the fit to the parametric model
\eref{rm} for the same three voxels. Although there is no restriction
on the form of the response function, the estimated response always
has the expected shape: start at zero, rapid upslope, a short plateau
and a nearly exponential downslope. The response function is well
characterized by using \eref{rm}, and gives robust estimates of the
kinetic parameters across the given ROI. However, the first part of
the response function is typically slightly underestimated and the
second part is slightly overestimated. A more flexible parametric
model might fit the response function better.

\begin{figure}[p]
\begin{center}
\begin{tabular}{ccccc}
& Patient~2 & Patient~3 & Patient~9\\
\rotatebox{90}{\hspace{-3cm}SSR parametric}&
\includegraphics[angle=270,scale=.27,viewport=75 230 475 640]{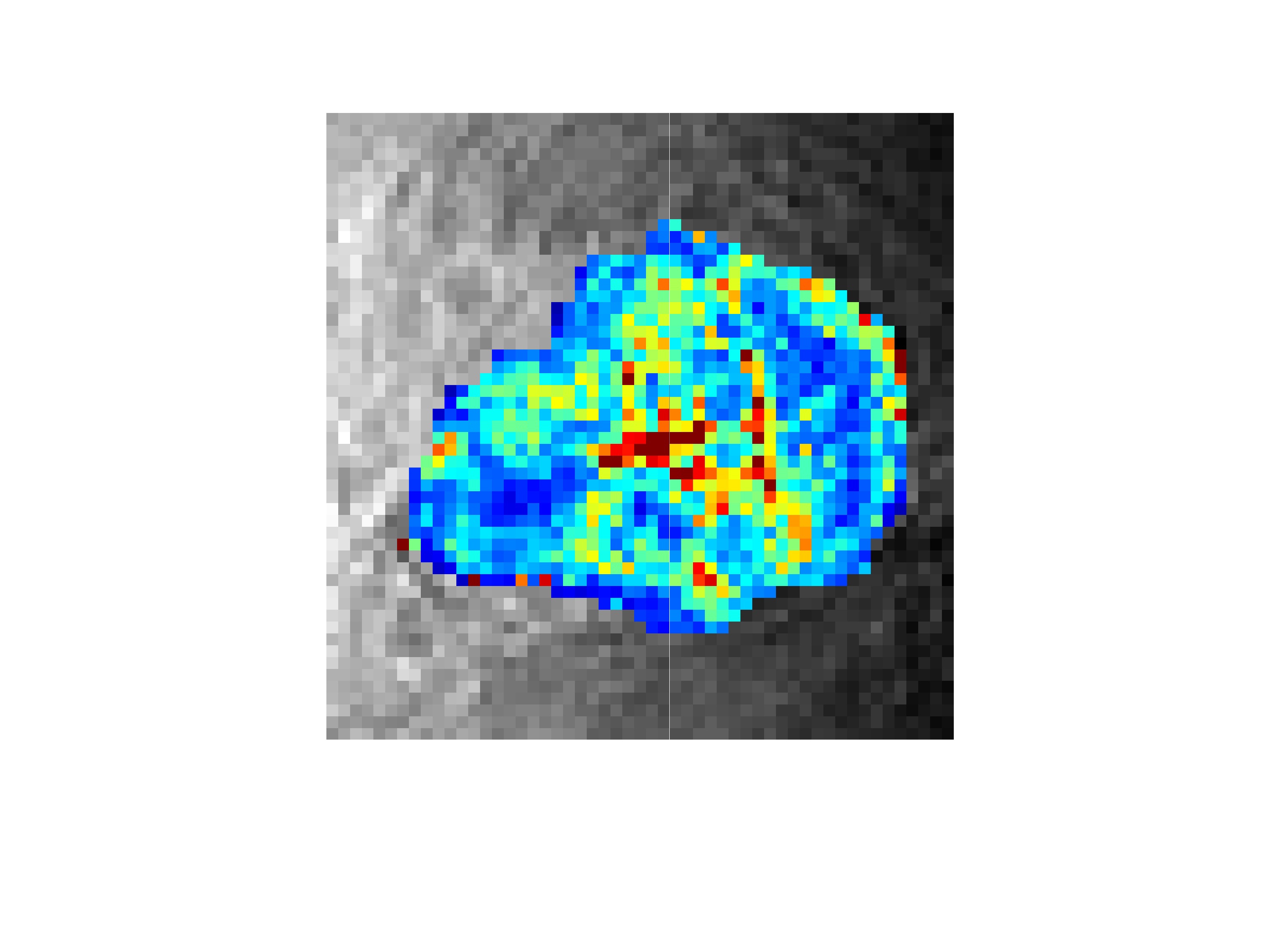}&
\includegraphics[angle=270,scale=.27,viewport=75 230 475 640]{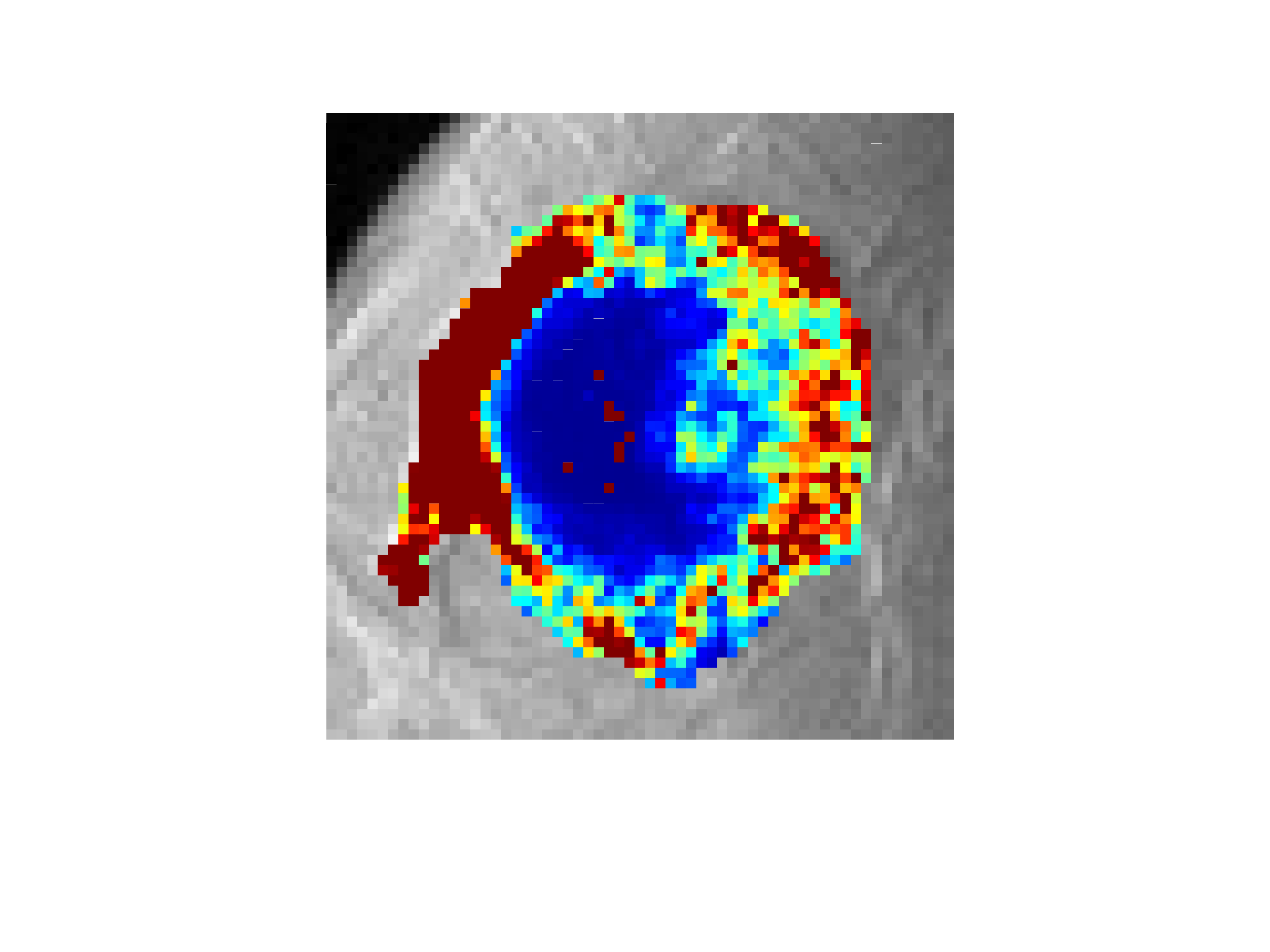}&
\includegraphics[angle=270,scale=.27,viewport=75 230 475 640]{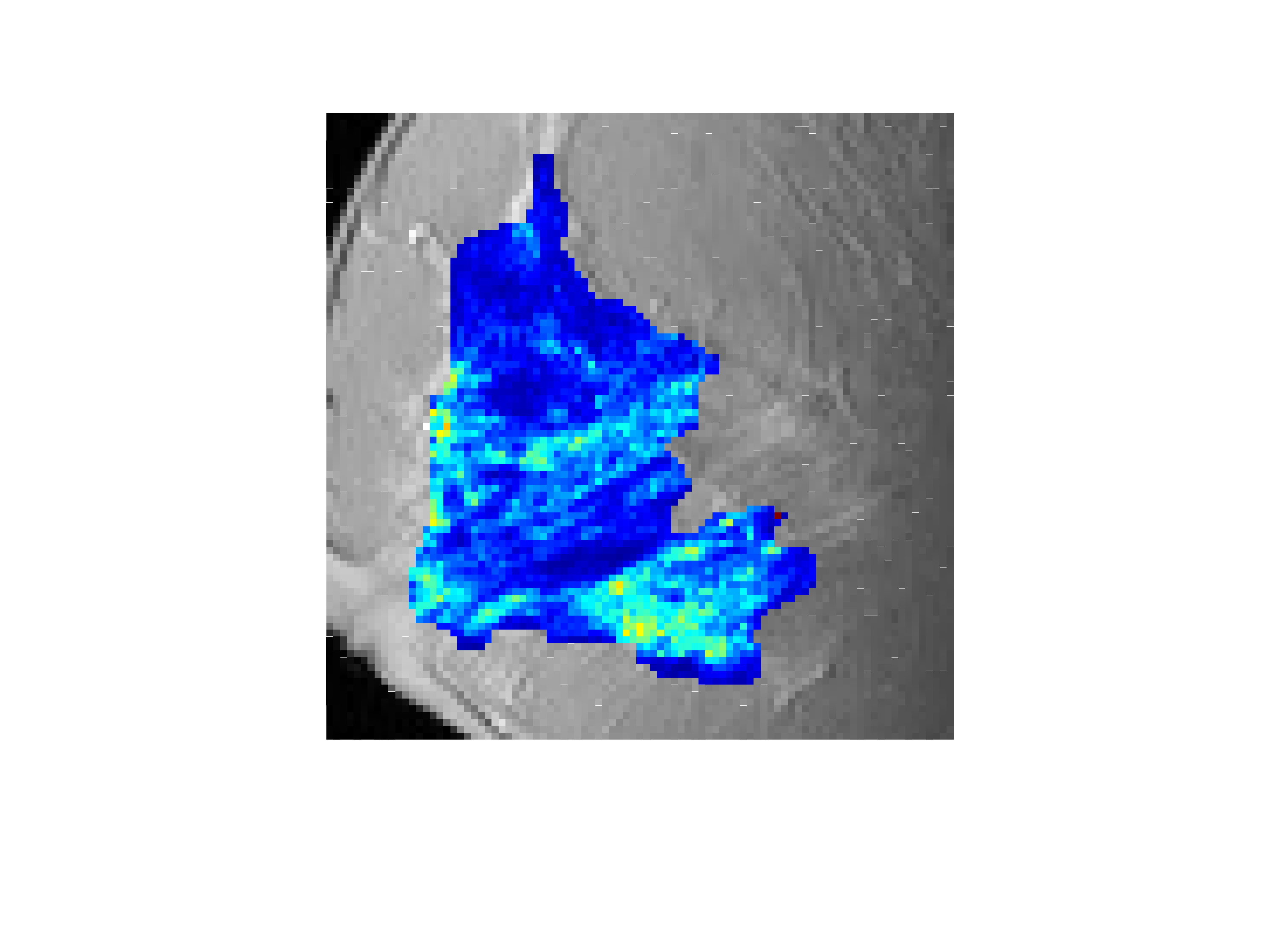}&\\
\rotatebox{90}{\hspace{-3.5cm}SSR semi-parametric}&
\includegraphics[angle=270,scale=.27,viewport=75 230 475 640]{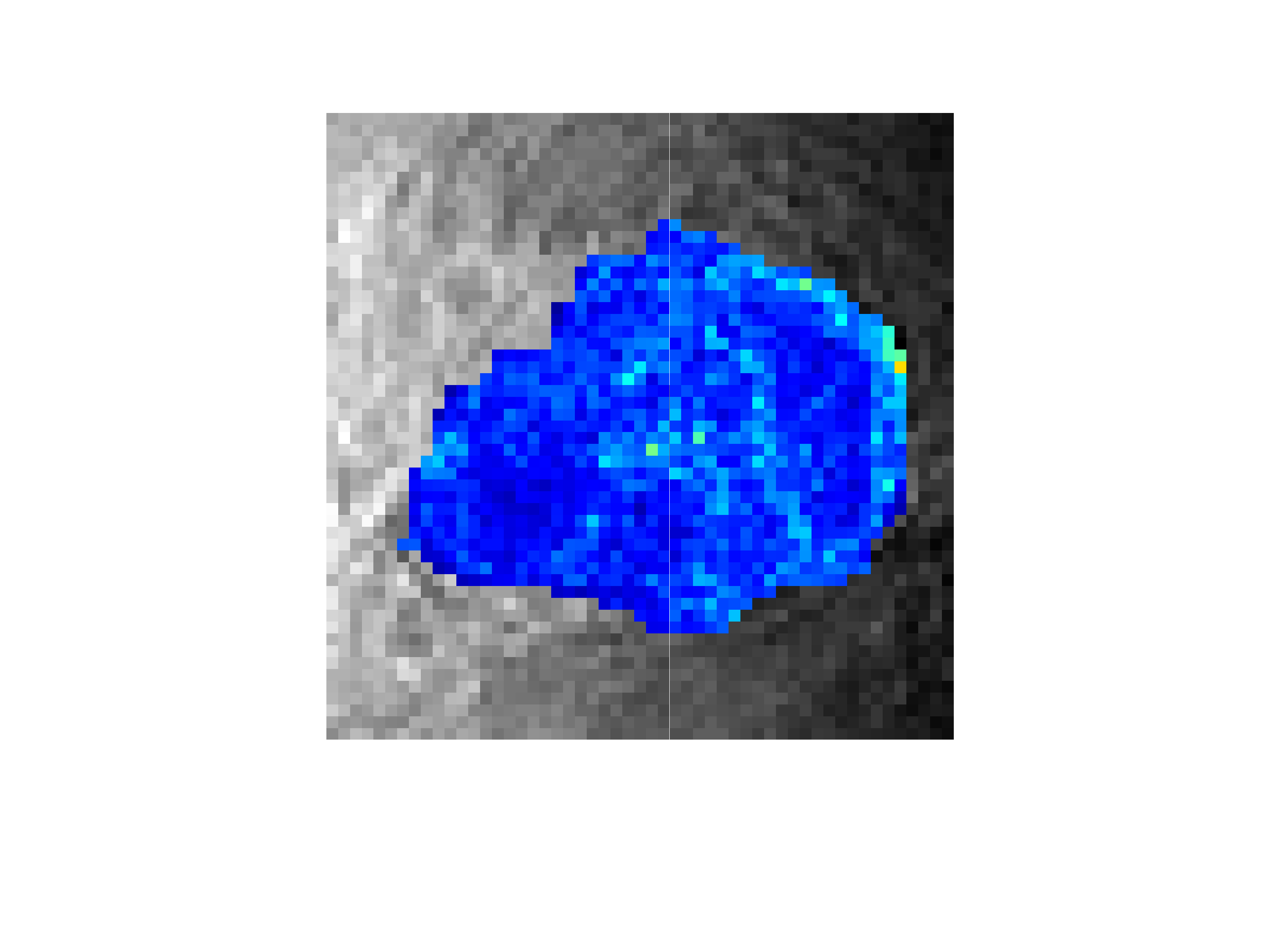}&
\includegraphics[angle=270,scale=.27,viewport=75 230 475 640]{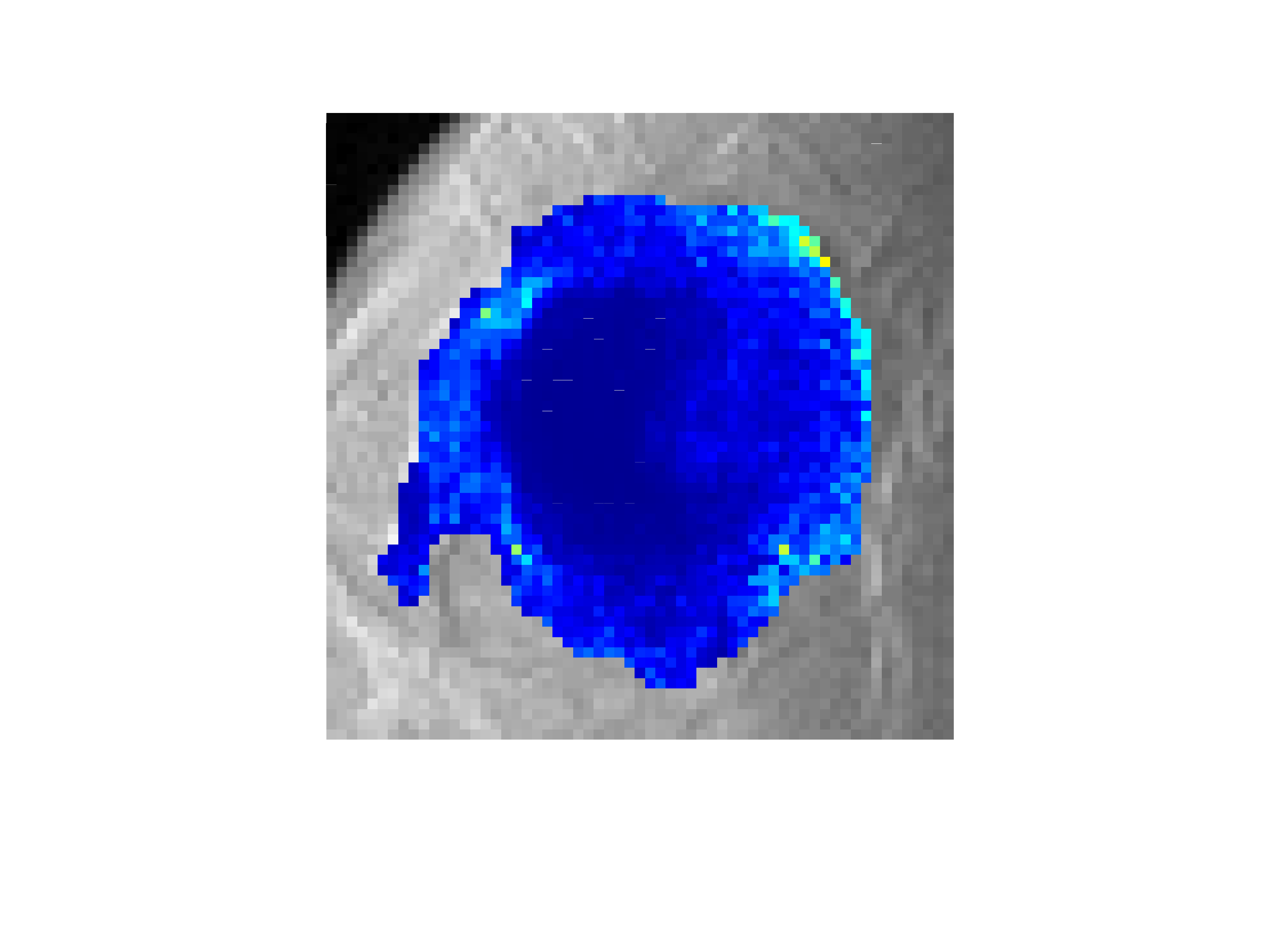}&
\includegraphics[angle=270,scale=.27,viewport=75 230 475 640]{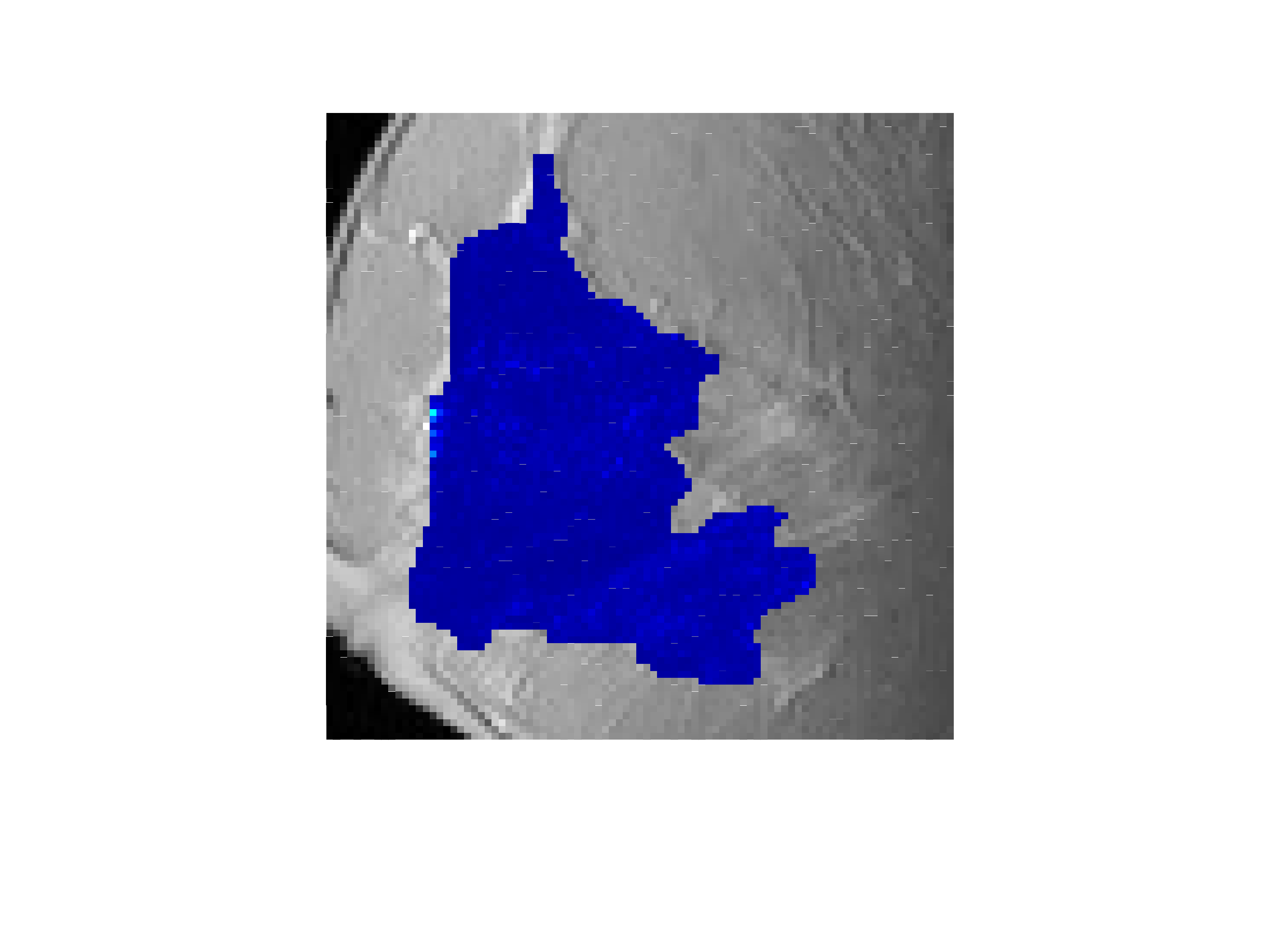}&\\
\rotatebox{90}{\hspace{-3cm}$\ktrans$ parametric}&
\includegraphics[angle=270,scale=.27,viewport=75 230 475 640]{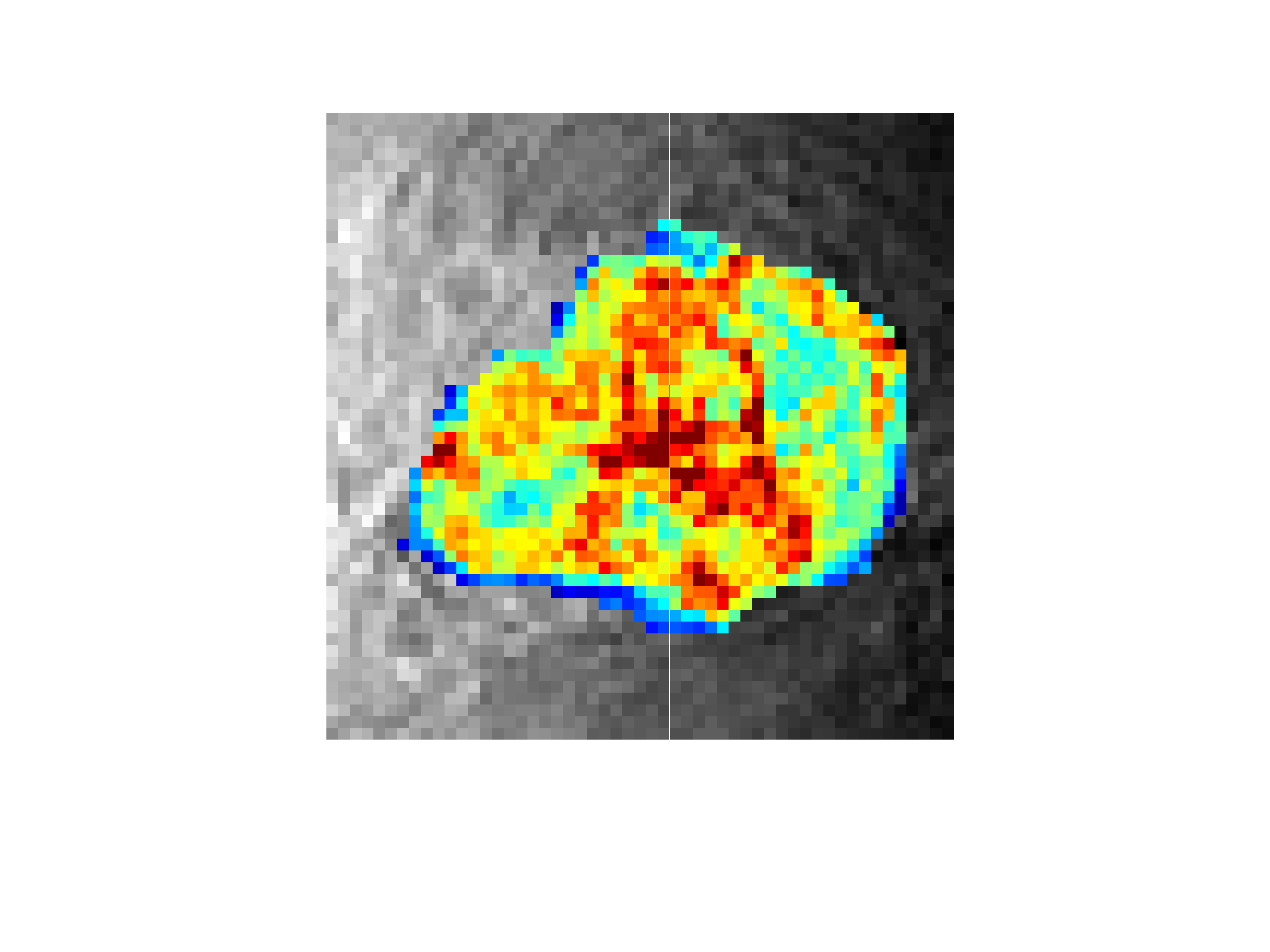}&
\includegraphics[angle=270,scale=.27,viewport=75 230 475 640]{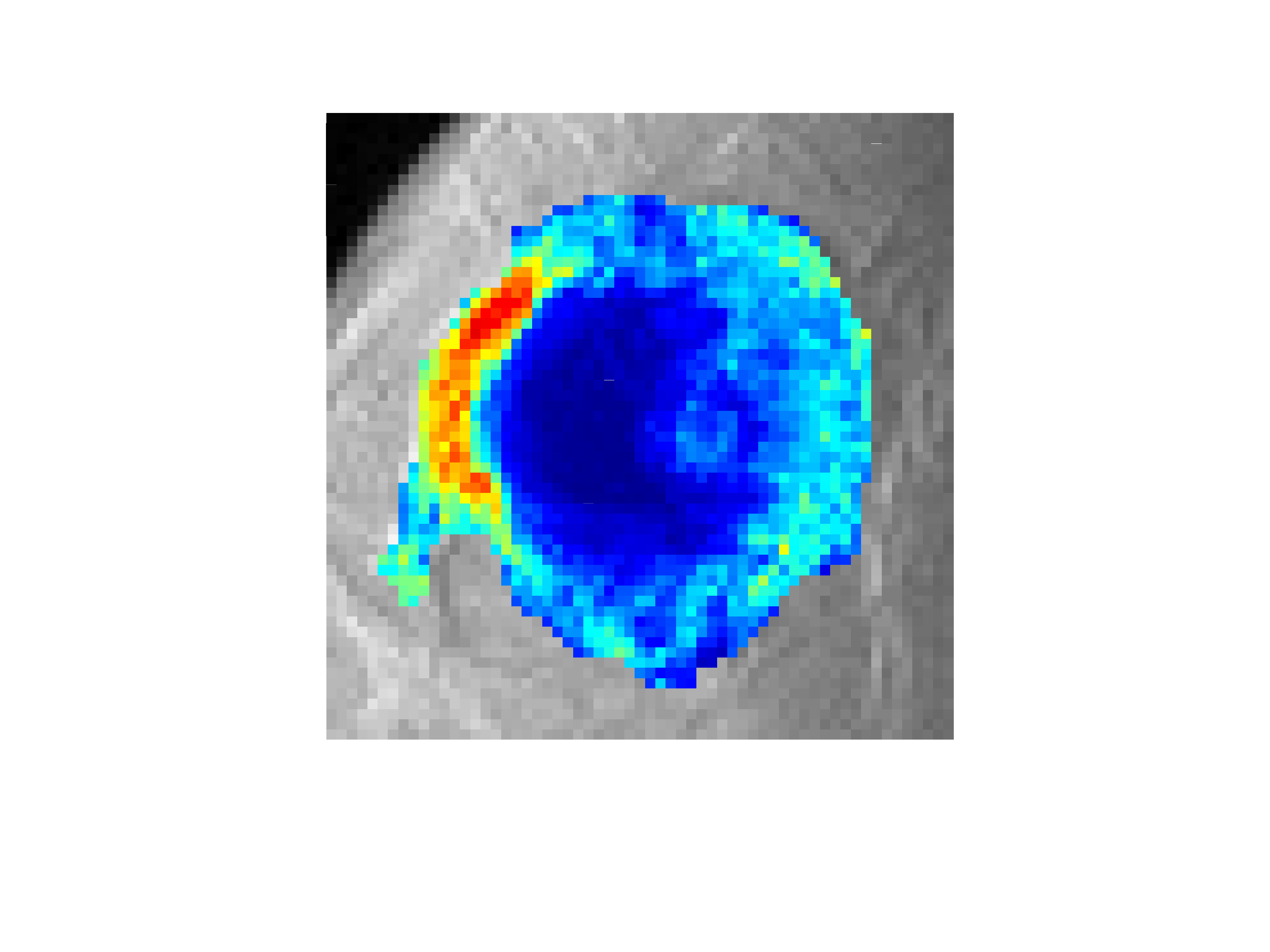}&
\includegraphics[angle=270,scale=.27,viewport=75 230 475 640]{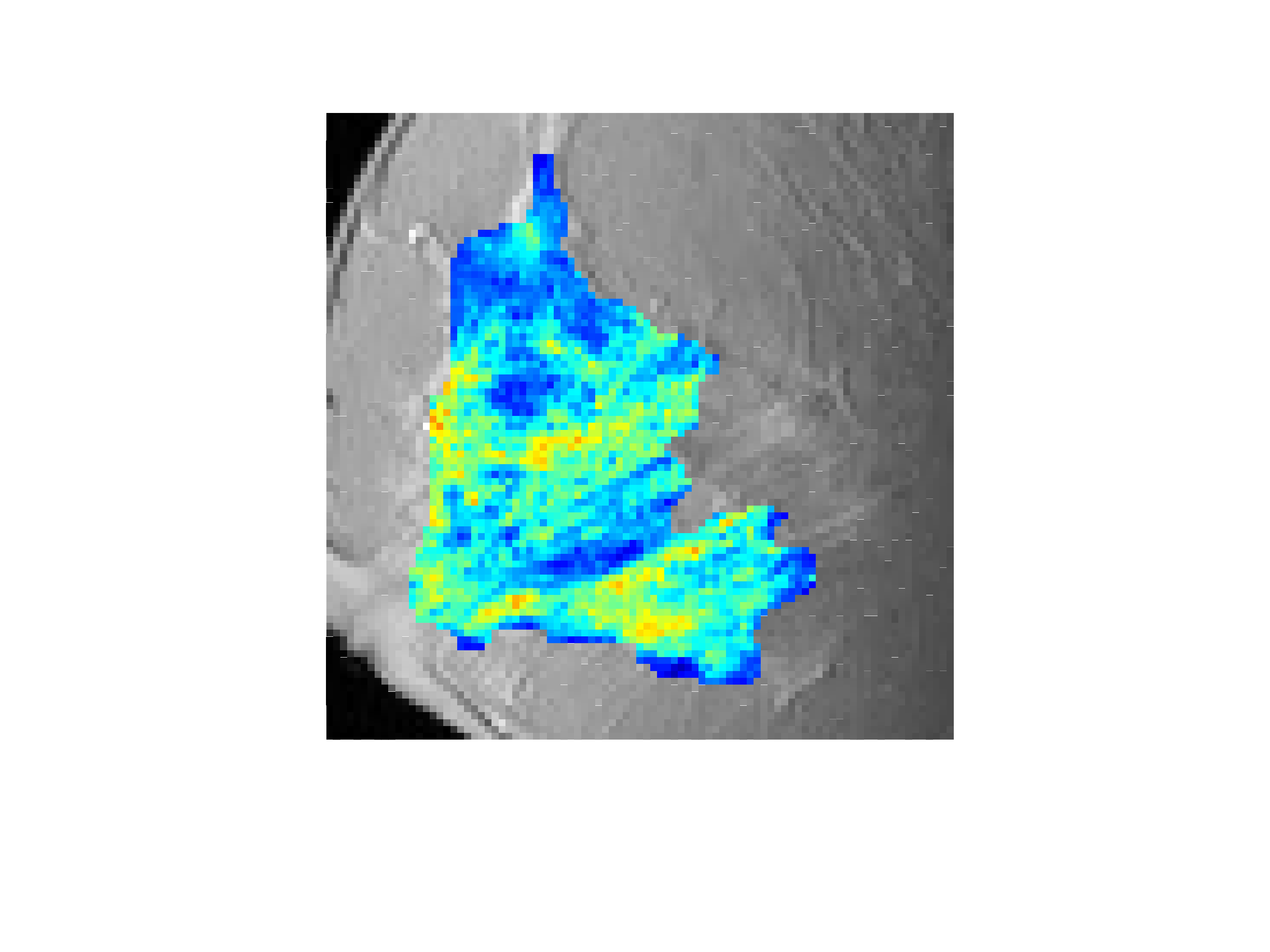}&\\
\rotatebox{90}{\hspace{-3.5cm}$\ktrans$ semi-parametric}&
\includegraphics[angle=270,scale=.27,viewport=75 230 475 640]{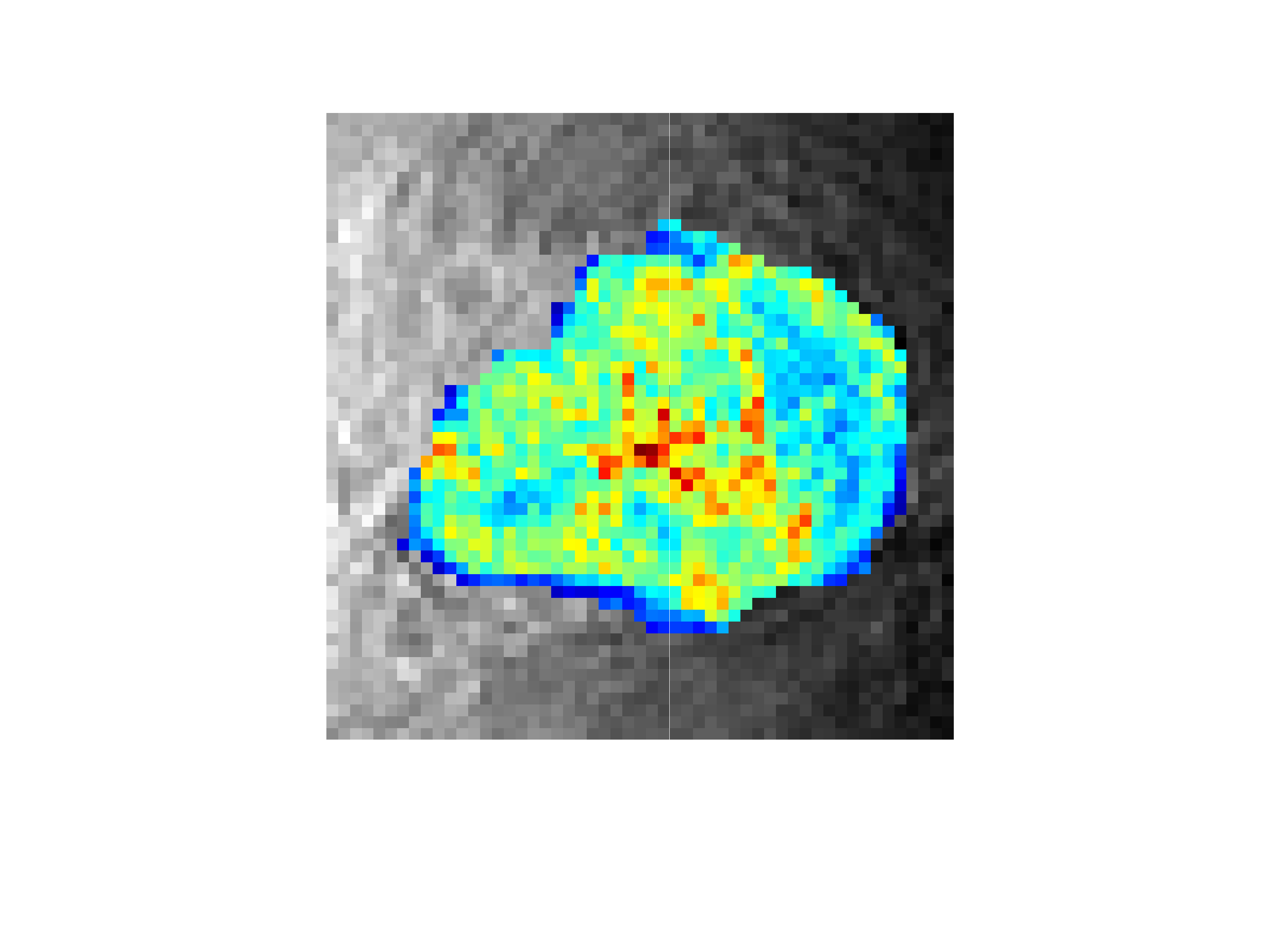}&
\includegraphics[angle=270,scale=.27,viewport=75 230 475 640]{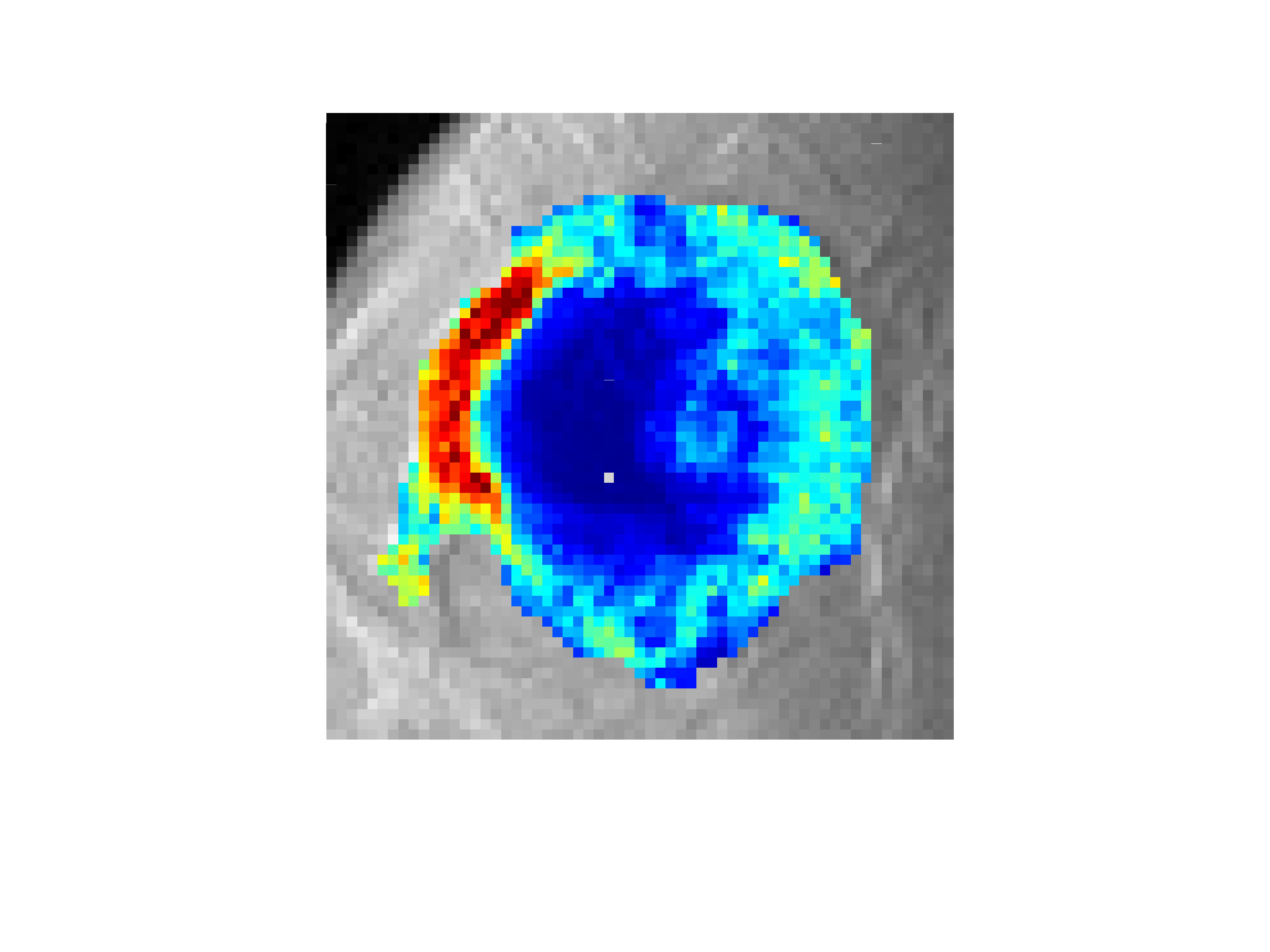}&
\includegraphics[angle=270,scale=.27,viewport=75 230 475 640]{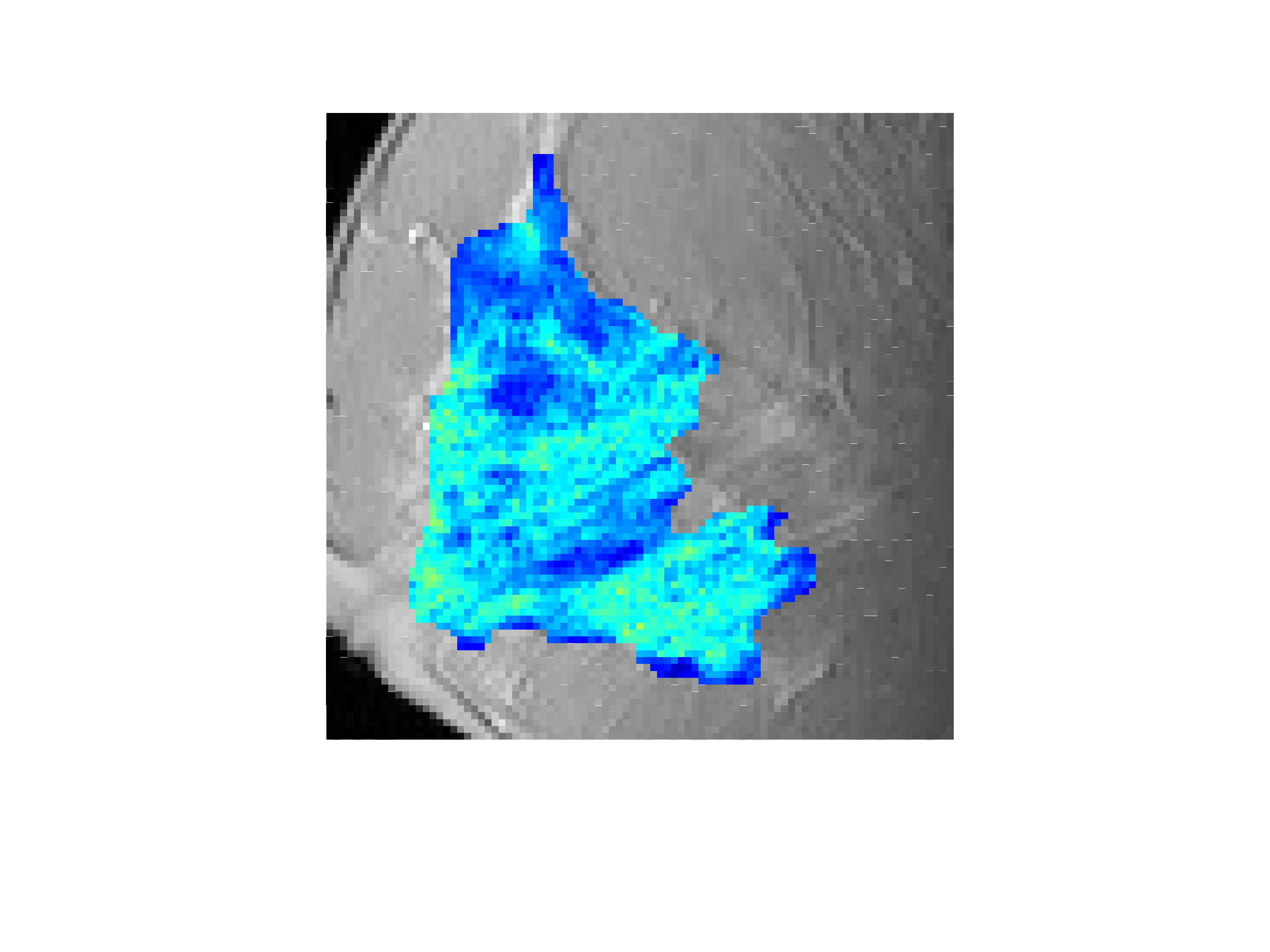}&\\
\rotatebox{90}{\hspace{-3.5cm}standard error $\ktrans$}&
\includegraphics[angle=270,scale=.27,viewport=75 230 475 640]{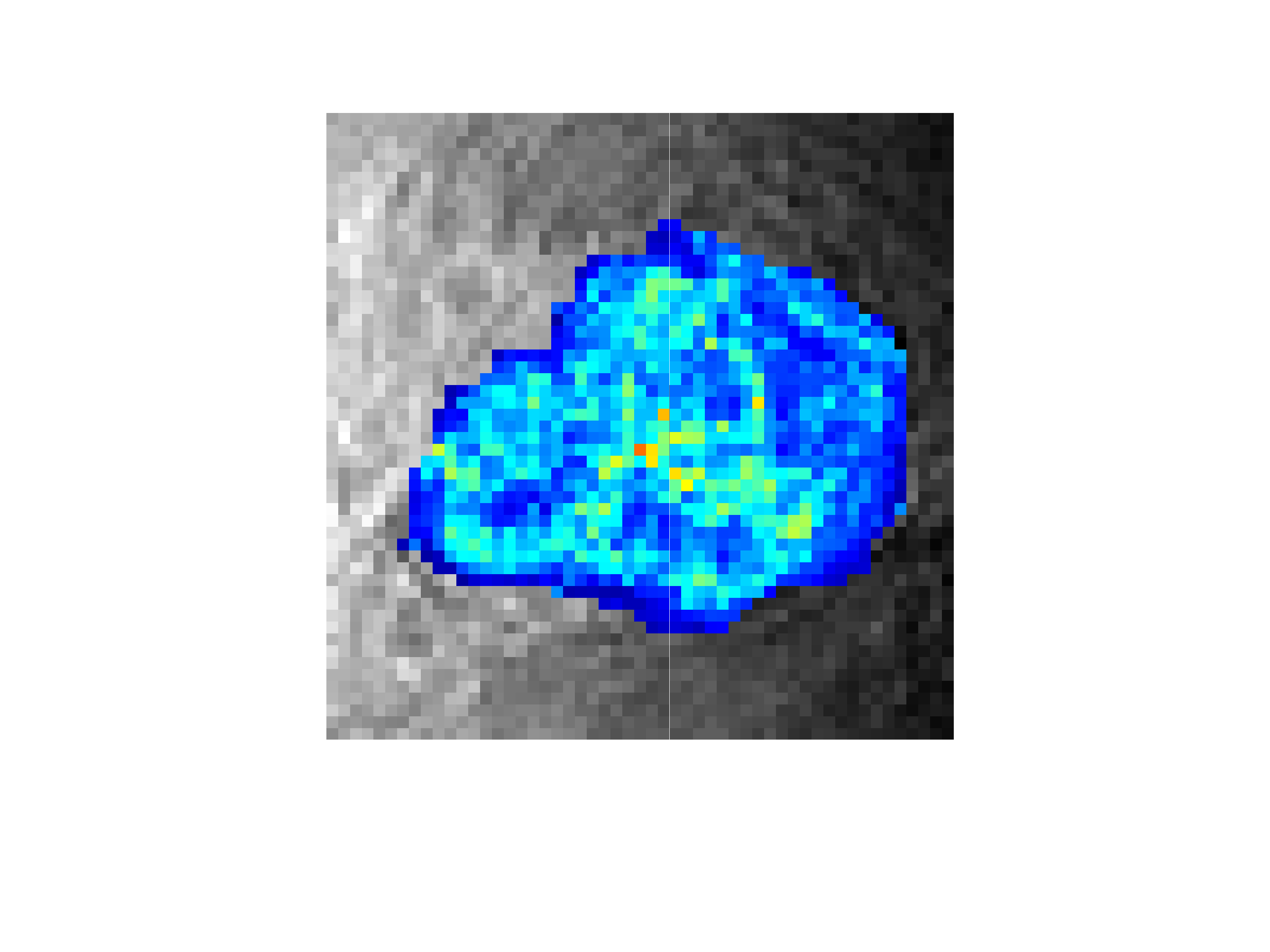}&
\includegraphics[angle=270,scale=.27,viewport=75 230 475 640]{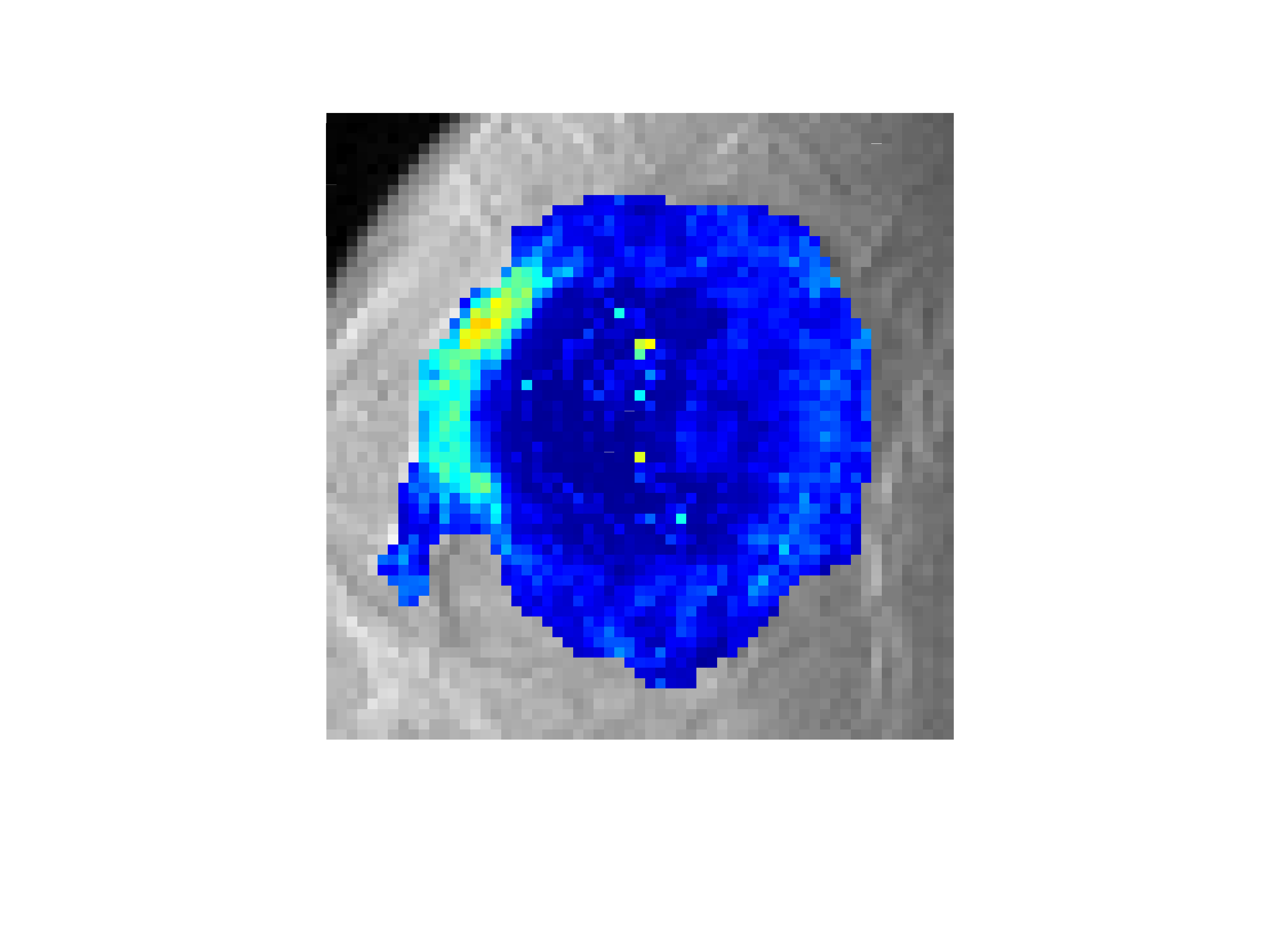}&
\includegraphics[angle=270,scale=.27,viewport=75 230 475 640]{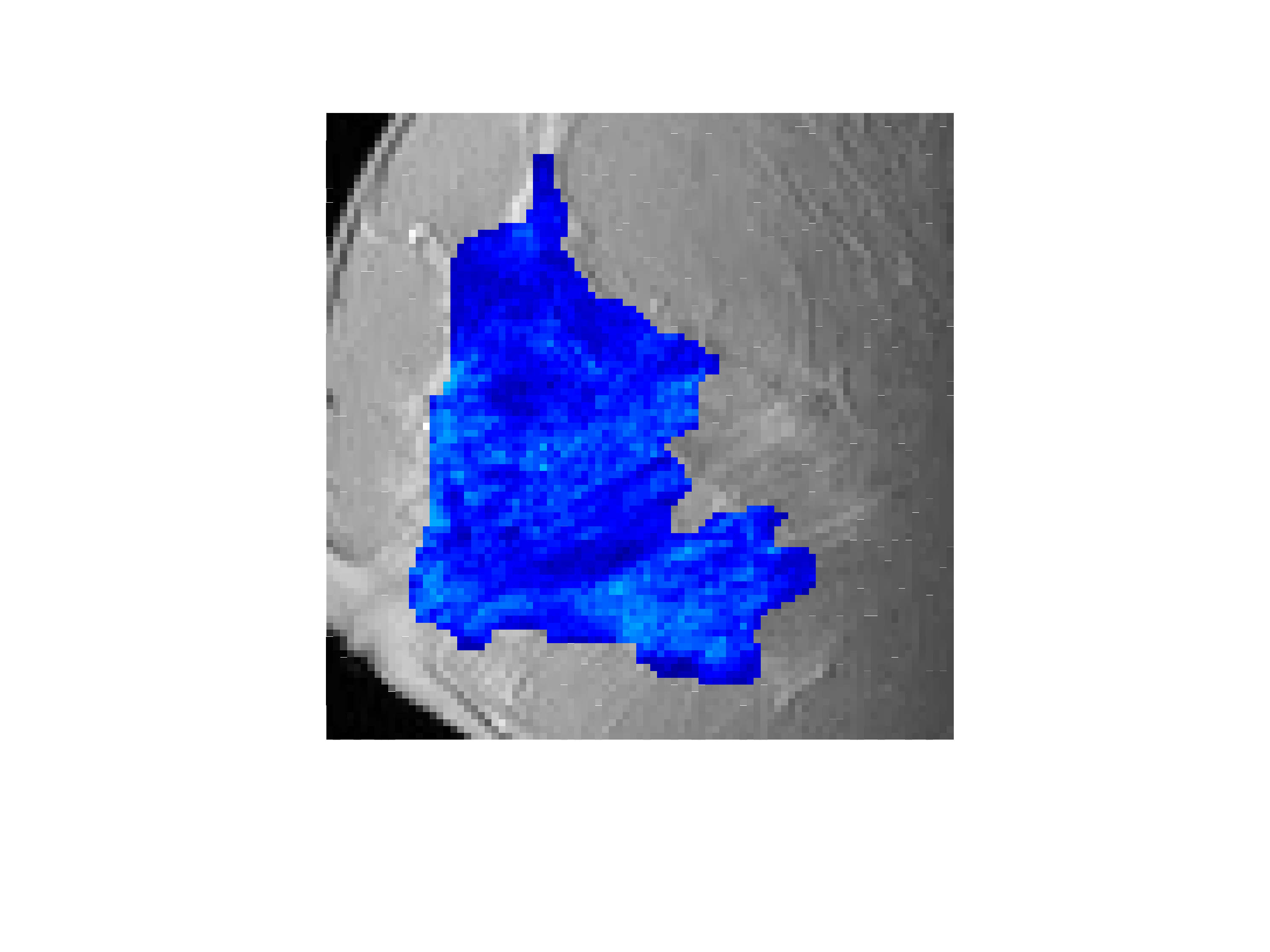}&\\
\\
\end{tabular}
\caption{Parameter maps for the first scan of patients 2, 3 and
  9. From top to bottom: Sum of squared residuals (SSR) map for
  reference parametric technique; SSR map for proposed semi-parametric
  technique; $\ktrans$ parameter map estimated with reference
  parametric technique; $\ktrans$ parameter map estimated with
  proposed semi-parametric technique standard error of $\ktrans$
  estimated from proposed semi-parametric technique.}
\label{fig:parametermaps2}
\end{center}
\end{figure}

Fig.~\ref{fig:parametermaps2} (rows 1 and 2) depicts the sum of
squared residuals (SSR) in the ROI for the semi-parametric technique
and the reference parametric method for the pre-treatment scans of
three patients.  SSRs are clearly reduced over the whole area of the
tumor, especially in areas with high $\ktrans$ values, \ie, in areas
with fast blood flow. Therefore, $\ktrans$ estimates are clearly
different for both methods, parameter maps are shown in
Fig.~\ref{fig:parametermaps2} (row 3 and 4). There is, however, no
general trend for under- or overestimation with different methods. For
example, for subject 9, $\ktrans$ estimates are higher with the
parametric method, as the upslope is overestimated (see also
Fig.~\ref{fig:fit}).  On the other hand, for patient 3, $\ktrans$
estimates are to low by using the parametric technique due to a wrong
onset time.  Again, the Bayesian method allows us to access the
precision of the estimated kinetic
parameters. Fig.~\ref{fig:parametermaps2} (row 5) depicts the standard
error of the $\ktrans$ parameter computed with the semi-parametric
technique.  Estimation errors are higher with higher values, but they
are generally relatively small. For most voxels, the relative error,
\ie, standard error divided by estimated parameter, does not exceed
20\%.

\section{Conclusion}

In this paper, we have introduced a semi-parametric technique based on
Bayesian P-spline models for quantifying CTCs obtained with DCE-MRI.
Compared with parametric models, the proposed semi-parametric
technique provides a superior fit to the observed concentration time
curves.  In particular, the proposed semi-parametric method captures
the upslope of the time series accurately, which is important for an
accurate fit of the CTCs in DCE-MRI and proper calculation of
$\ktrans$.

Results from the simulated data show a clear improvement in both time
curve fit and parameter estimation. In addition, the Bayesian method
provides information about the precision of the estimation, including
standard errors or credible intervals. For \textit{in vivo}
validation, the fit of the contrast agent concentration with the
proposed semi-parametric technique is superior to the parametric
method. Estimates of kinetic parameters for both techniques are
different where the fit of the parametric technique is poor,
suggesting that the estimation of kinetic parameters with the proposed
semi-parametric technique is more accurate in these areas.  The exact
assessment of the kinetics is clinically important as changes in
respect to the response function and the kinetic parameters can be
subtle, especially for drugs that cause multiple effect on tumor
vasculature, such as combinations of antiangiogenic drugs
\cite{padhani05b}.

Non-linear parametric models are typically difficult to estimate due
to the convergence issues and problems in specifying consistent
starting values. Bayesian non-linear regression can overcome these
convergence problems, but at a cost of computational time
\cite{schmid05m}.  The proposed semi-parametric method provides a
reliable and practical way of circumventing these problems.  With the
proposed technique, computation only includes sampling from standard
distributions (see section \ref{fullcond}) which can be done
efficiently \cite{rue01} and gives good mixing of the MCMC kernel.

In contrast to classical approaches \cite{jeroschherold02}, Bayesian
P-splines allow simultaneous estimation of model and smoothing
parameters. Adaptive smoothing parameters can be obtained and are
important to model the sharp changes in the dynamic series of the
contrast concentration.

One unique feature of the paper is the estimation of the onset of the
CTC based on Bayesian inference. Onset time can be determined for the
whole region of interest or on a more local level.  It is an important
parameter in quantitative DCE-MRI models as it has great influence on
other parameters in the kinetic model and an incorrect onset time can
lead to a strong bias in parameter estimates \cite{orton07}. Onset
time is also an important clinical index for characterizing suspicious
breast lesions \cite{boetes94}, although in this study they are not
evident for the patients studied.

The proposed method allows a direct assessment of the response
function, \ie, the actual flow in the tissue.  Parameters of interest
may be estimated by fitting a non-linear model to the response
function. Smoothing via P-splines provides an effective way of error
reduction and deconvolution of the arterial input function.  The
proposed technique also allows the quantification of errors both in
fitting the observed data and in estimating kinetic parameters.  Thus
far, the estimation error in DCE-MRI models is rarely discussed, the
proposed technique can contribute to the evaluation of the quality of
DCE-MRI scans.  Results from the semi-parametric technique can
therefore serve as a supporting tool for quality control. When
choosing a pixel, the CTC would be displayed interactively along with
the semi-parametric fit, the CI, and the estimated onset time. This
would not only help to reassess the quality of the data, but also make
it easier to understand what is actually happening physiologically.

\section*{Acknowledgments}

Support for Volker Schmid was supported by a platform grant
(GR/T06735/01) from the UK Engineering and Physical Sciences Research
Council (EPSRC). He was also partially financed through a research
grant from GlaxoSmithKline.  We are grateful to David Buckley at
Imaging Science and Biomedical Engineering, University of Manchester,
UK for providing the simulated data.  The clinical data were
graciously provided by Anwar Padhani and Jane Taylor at Paul
Strickland Scanner Centre, Mount Vernon Hospital, Northwood, UK.

\bibliographystyle{IEEEtran}
\bibliography{../mripapers,../badnews}

\onecolumn

\begin{table*}[bt]
\begin{center}
\caption{Values of the kinetic parameters describing the
behavior of MMID4 used to simulate 13 different contrast
concentration time curves representative of a breast tumor.
\label{tab:simvalues}
}
\begin{tabular}{c||c|cccc|cccc|cccc}
Exp.~& Baseline  & 2&3&4&5&6&7&8&9&10&11&12&13 \\\hline
$F_p$ & 0.57 & 0.17 & 0.37 & 0.77 & 0.97 & \multicolumn{4}{c|}{0.57} & \multicolumn{4}{c}{0.57} \\
$v_p$ & 0.06 & \multicolumn{4}{c|}{0.06} &$10^{-4}$& 0.03& 0.09& 0.12 &\multicolumn{4}{c}{0.06} \\
$PS$ & 0.33 &  \multicolumn{4}{c|}{0.33} &  \multicolumn{4}{c|}{0.33} & 0.01& 0.17& 0.49& 0.65\\
$v_e$ & 0.45 & \multicolumn{4}{c|}{0.45} &  \multicolumn{4}{c|}{0.45} & \multicolumn{4}{c}{0.45} \\
\end{tabular}
\end{center}
\end{table*}

\begin{table*}[!tb]
\begin{center}
\caption{True and estimated values of $\ktrans$ with 95\% CI.\label{fig:ktranserrorbar}}
\begin{tabular}{l||r|r|r|r|r|r|r|r|r|r|r|r|r}
Experiment & 1 & 2 & 3 & 4 & 5 & 6 & 7 & 8 & 9 & 10 & 11 & 12 & 13 \\\hline
True $\ktrans$ &  0.251 & 0.146 & 0.218 & 0.268 & 0.280  & 0.233 & 0.233 & 0.233 & 0.233 & 0.010 & 0.147 & 0.323 & 0.388\\\hline
2.5\% quantile $\ktrans$   & 0.208 & 0.128 & 0.201 & 0.236 & 0.242  & 0.181 & 0.220 & 0.224 & 0.229 & 0.004 & 0.129 & 0.293 & 0.347\\
Median posterior $\ktrans$ & 0.245 & 0.138 & 0.218 & 0.263 & 0.284  & 0.200 & 0.237 & 0.281 & 0.282 & 0.005 & 0.147 & 0.320 & 0.376\\
97.5\% quantile $\ktrans$  & 0.267 & 0.167 & 0.240 & 0.309 & 0.376  & 0.239 & 0.256 & 0.334 & 0.372 & 0.060 & 0.183 & 0.356 & 0.415\\
\end{tabular}
\end{center}
\end{table*}

\begin{table*}[!tb]
\begin{center}
\caption{Sum of squared residuals (SSR) between data and
  fitted function for parametric and semi-parametric method, averaged
  over all voxels in the ROI. \label{tab:vglpatients}}
\begin{tabular}{l||r|r|r|r|r|r|r|r|r|r|r|r}
  1st scan, subject & 1 & 2 & 3 & 4 & 5 & 6 \\\hline
  parametric & 0.0682 & 0.1579 & 0.2209 & 0.1825 & 0.6101 & 0.4382 \\
  semi-parametric & 0.0217 & 0.0575 & 0.0341 & 0.183 & 0.0398 & 0.0333\\\hline
   & 7 & 8 & 9 & 10 & 11 & 12 \\\hline
  parametric & 0.0723 & 0.0366 & 0.0798 & 0.0568 & 0.1420 & 0.0552\\
  semi-parametric & 0.0242 & 0.0068 & 0.104 & 0.0077 & 0.0131 & 0.0234 \\\hline\hline
  2nd scan, subject & 1 & 2 & 3 & 4 & 5 & 6 \\\hline
  parametric & 0.0457 & 0.0691 & 0.0403 & 0.8665 & 0.6469 & 0.5804\\
  semi-parametric & 0.0123 & 0.0328 & 0.0122 & 0.0484 & 0.0386 & 0.0334 \\\hline
   & 7 & 8 & 9 & 10 & 11 & 12 \\\hline
  parametric & 0.0998 & 0.0525 & 0.0237 & 0.0544 & 0.0225 & 0.0289\\
  semi-parametric & 0.0209 & 0.0102 & 0.0071 & 0.0205 & 0.0072 & 0.0168
\end{tabular}
\end{center}
\end{table*}

\begin{table*}[!tb]
\begin{center}
\caption{Estimated contrast concentration enhancement onset time in seconds for all 24 scans. Protocol specified that the tracer was injected after 49.40 seconds.\label{tab:t0}}
\begin{tabular}{l||r|r|r|r|r|r|r|r|r|r|r|r}
Subject & 1 & 2 & 3 & 4 & 5 & 6 & 7 & 8 & 9 & 10 & 11 & 12\\\hline
1st scan & 55.19 & 54.14 & 69.09 & 90.20 & 60.04 & 57.63 & 52.56 & 54.95 & 55.85 & 56.72 & 47.48 & 50.34 \\
2nd scan & 55.09 & 56.58 & 87.09 & 56.58 & 56.71 & 55.31 & 53.10 & 55.46 & 53.33 & 53.96 & 54.06 & 49.59
\end{tabular}
\end{center}
\end{table*}

\newpage

\end{document}